\documentclass[%
 reprint,
 amsmath,amssymb,
 aps,
floatfix,
]{revtex4-1}

\usepackage{graphicx}% Include figure files
\usepackage{dcolumn}% Align table columns on decimal point
\usepackage{bm}% bold math
\usepackage{siunitx}
\usepackage{xcolor}

\newcommand{\Df}{\left\lvert \Delta V \right\rvert}
\newcommand{\er}{\varepsilon_f\varepsilon_0}
\newcommand{\sd}{\omega}
\graphicspath{{figs/}}

\begin{document}

\title{Dielectric breakdown by electric-field induced phase separation}

% Authors
\author{Dimitrios Fraggedakis$^1$}
% \thanks{aka dfrag - email: dimfraged@gmail.com}
\thanks{email: dimfraged@gmail.com}
\author{Mohammad  Mirzadeh$^1$}
\author{Tingtao Zhou$^2$}
\author{Martin Z. Bazant$^{1,3}$}
\thanks{email: bazant@mit.edu}
\affiliation{
   $^1$Department of Chemical Engineering,
   Massachusetts Institute of Technology,
   Cambridge, MA 02139 USA
   }
\affiliation{
  $^2$ Department of Physics,
   Massachusetts Institute of Technology,
   Cambridge, MA 02139 USA
   }
 \affiliation{
  $^3$Department of Mathematics,
   Massachusetts Institute of Technology,
   Cambridge, MA 02139 USA
   }
% \affiliation{
%   Massachussets Institute of Technology,\\
%   Cambridge, MA 02139 USA
% }

\date{\today}% It is always \today, today,

\begin{abstract}
The control of the dielectric and conductive properties of device-level systems is important for increasing the efficiency of energy- and information-related technologies. In some cases, such as neuromorphic computing, it is desirable to increase the conductivity of an initially insulating medium by several orders of magnitude, resulting in effective dielectric breakdown. Here, we show that by tuning the value of the applied electric field in systems { with variable permittivity and electric conductivity}, e.g. ion intercalation materials, we can vary the device-level electrical conductivity by orders of magnitude. We attribute this behavior to the formation of filament-like conductive domains that percolate throughout the system, { which form only when the electric conductivity depends on the concentration}. We conclude by discussing the applicability of our results in neuromorphic computing devices and Li-ion batteries. 
\end{abstract}

\maketitle

%\tableofcontents

\section{\label{sec:introduction}Introduction}

Phase separating materials play a key role in several applications related to energy harvesting and storage, as well as information storage and processing. Some characteristic examples are Li-ion batteries~\cite{lim2016origin}, phase change~\cite{ielmini2018memory} and redox~\cite{fuller2017li} memristive devices~\cite{waser2019introduction}, alloy catalysts~\cite{zafeiratos2012alloys} and self-organized surface nanoreactors~\cite{hildebrand1999self}. In some cases, phase separation is desirable as one can increase the efficiency of the system (e.g. increase catalytic activity~\cite{Mikhailov2009}, change of electric~\cite{nadkarni2019modeling,gonzalez2020lithium} and/or thermal~\cite{lu2020bi} conductivity). However, there are other cases where phase separation degrades the performance of a device resulting in decreased lifetime (e.g. fracture of secondary electrode particles in Li-ion batteries~\cite{t2016phase,di2015diffusion} that decreases the available active material, delamination at electrode-electrolyte interface in all-solid-state batteries~\cite{Koerver2018} resulting in loss of contact of active material with the electrode). Hence, it is crucial to find ways that can actively control the occurrence and/or suppression of phase separation, which can help us increase the efficiency of existing technologies, as well as open new possibilities on exploiting physical phenomena for new applications.

Phase separation can be described as a form of instability. For example, a homogeneous binary mixture is thermodynamically unstable when its average concentration lies inside the spinodal region~\cite{kondepudi2014modern}. In this situation, any infinitesimal perturbation on the concentration field would evolve in time, making the system to form domains of the two phases~\cite{balluffi2005kinetics}. In general, there have been several efforts to control or induce the formation of instabilities in equilibrium and non-equilibrium systems. Some examples are: 1) control of viscous fingering through the application of electric fields~\cite{mirzadeh2017electrokinetic,gao2019active}, 2) stabilization of thermodynamically unstable mixtures used in Li-ion batteries, such as LiFePO$_4$, using non-equilibrium driving forces, galvanostatic conditions~\cite{bai2011suppression,lim2016origin,bazant2017thermodynamic,fraggedakis2020tuning} 3) destabilization of homogeneous polymeric, colloidal, electrolyte and glass mixtures under the application of electric field~\cite{carmack2018tuning,gu2000effect,gu2000effect,thornburg1972electric,hori2007phase,aranson2002phase,Tsori2007,Tsori2004,wang2008effects}. In the present work, we are interested in understanding the control of phase separation of mixtures through electric fields and its impact on the transport properties of the system, e.g. change of electric conductivity after phase separation occurs.

In many dielectric mixtures that phase separate under electric fields, e.g. colloids~\cite{khusid1996effects,khusid1999phase,kumar2004combined,Johnson2004}, polymers~\cite{Liedel2012}, amorphous solids~\cite{thornburg1972electric}, electrolytes~\cite{Tsori2004}, the electric permittivity depends on the corresponding species concentration~\cite{tomlinson1972spinodal}. The effect of this dependence can be understood in the simple case of a binary mixture, where the applied electric field contributes to the total chemical potential $\mu$, i.e. $\mu_{\mathbf{E}}\sim\partial_c\varepsilon\left\lvert\mathbf{E}\right\rvert^2$~\cite{bazant2013theory}. { Therefore, a combination of a nonlinear concentration-dependent permittivity combined with high electric fields can alter the free energy to change the miscibility gap and spinodal region.} 

The effect of the electric field has mainly to do with the thermodynamics stability, however, the formed phase morphologies, e.g. filament-like structures~\cite{kupershtokh2006anisotropic,liedel2012beyond}, can greatly affect the transport properties on the macroscopic level, e.g. from low to high electric conductance and vice versa,~\cite{gonzalez2020lithium}, leading to phenomena that resemble dielectric breakdown~\cite{sze2006physics,o1973theory}. { Previous studies have focused identifying the conditions for dielectric breakdown due to phase separation solely based on thermodynamics~\cite{thornburg1972electric,gonzalez2020lithium,khusid1996effects,khusid1999phase,kumar2004combined,Johnson2004}, and few studies have discussed the formation and dynamics of the conductive filaments~\cite{kupershtokh2006anisotropic}, which is crucial for technological applications~\cite{ielmini2018memory}.} There are several solid-state materials that form conductive and insulating domains when they phase separate. For example, most of the commercial Li-ion intercalation materials, i.e. Li$_x$CoO$_2$~\cite{milewska2014nature,nadkarni2019modeling}, Li$_x$Ni$_{1/3}$Co$_{1/3}$Mn$_{1/3}$O$_2$~\cite{amin2016characterization}, Li$_{4+3x}$Ti$_5$O$_{12}$~\cite{verde2016elucidating,young2013electronic}, share this property as they undergo a metal-to-insulator transition (MIT), along with an ion concentration-dependent permittivity. This combination of characteristics makes them perfect candidates for studying the effect of electric fields on both the phase separation and the electric conduction, which can lead to dielectric breakdown.

The goal of the present work is to develop a simple phenomenological theory that describes dielectric breakdown due to electric-field induced phase separation. Based on a concentration-dependent electric permittivity, we show that a homogeneous stable solution can phase separate in two (or more) phases after a critical electric field is applied. Additionally, we consider the case where one of the phases is a metal and the other an insulator, { which translates in concentration-dependent electric conductivity}. When the initial concentration is such that the material is insulating, after phase separation occurs we find the system to conduct current like a metal. The transition from insulating to metallic behavior after electric field is applied corresponds to an effective dielectric breakdown, which is attributed to the formation of filament-like structures that span the entire domain. This phenomenon is related to the pioneering works by Goldhammer~\cite{goldhammer1913dispersion} \& Herzfeld~\cite{herzfeld1927atomic}, that link the changes in the electric permittivity with the species concentration to the metal-to-insulator transition. We relate our results to Li-ion intercalation materials, such as Li$_x$CoO$_2$ and Li$_{4+3x}$Ti$_5$O$_{12}$, and we discuss the implications of our theory for resistive switching and Li-ion battery applications. 

\section{\label{sec:modeling}Theory}

\begin{figure}[!ht]
    \centering
    \hspace{0.08in}\includegraphics[width=0.45\textwidth]{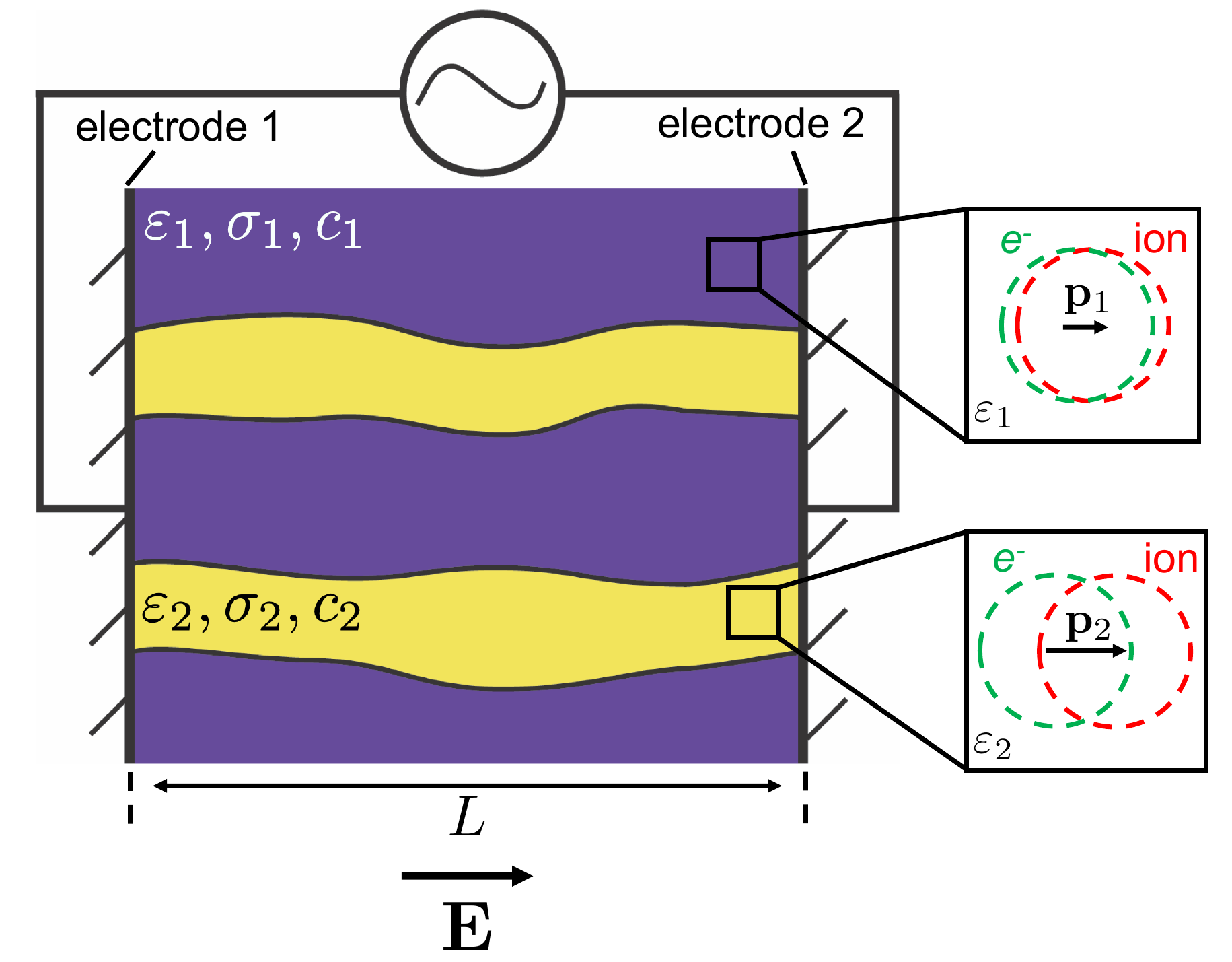}
    \caption{Schematic representation of a phase separating dielectric medium of length $L$ placed between two ion blocking electrodes. Each of the formed phases has different dielectric $\varepsilon$ and conductive $\sigma$ properties. The electric permittivity for each of the phases is described by a simple model based on two overlapping spheres of negative (green bound $e^-$) and positive (red ion) charges. The application of electric field $\mathbf{E}$ induce polarization $\mathbf{p}_i$ in each of the materials.}
    \label{fig:system}
\end{figure} 

We assume a phase separating dielectric medium placed between two blocking electrodes, as shown in Fig.~\ref{fig:system}. Purple and yellow show the two phases, respectively, that can be formed either because the mixture is thermodynamically unstable or due to the application of an electric field $\mathbf{E}$ that demixes the solid solution mixture. In the present picture, the two phases have different electric permittivities, $\varepsilon_1$ and $\varepsilon_2$, and electrical conductivities, $\sigma_1$ and $\sigma_2$. For simplicity we assume $\varepsilon_2>\varepsilon_1$ and $\sigma_2>\sigma_1$, respectively. Representative examples of such system are ion intercalation materials~\cite{manthiram2020reflection,Nitta2015,li2018fluid,nadkarni2018interplay}, which are thermodynamically unstable for a wide range of Li-ion fraction. Additionally, some of them undergo metal-to-insulator transition, e.g. Li$_x$CoO$_2$~\cite{milewska2014nature,nadkarni2019modeling} (LCO), Li$_{4+3x}$Ti$_5$O$_{12}$~\cite{verde2016elucidating,young2013electronic} (LTO), and Li$_x$TiO$_2$~\cite{de2017explaining,li2019low}, where one of the phases has much larger electrical conductivity compared to the other - the difference can be up to six orders of magnitude~\cite{sze2006physics}. In the following sections, we continue with the thermodynamics and transport theory that describes the electric-field induced phase separation and consequently the dielectric breakdown of the medium along the direction of the electric field.

\subsection{\label{subsec:thermo}Thermodynamics}
We are interested in modeling the dielectric phase separating medium shown in Fig.~\ref{fig:system}. The neutral species is described by the local fractional concentration $c=n/n_{max}$, where $n_{max}$ is the maximum species concentration in the medium. Under constant temperature and pressure, the Gibbs free energy of the system is~\cite{chaikin1995principles,bazant2013theory,gonzalez2020lithium,balluffi2005kinetics}
\begin{equation}
\label{eq:free_energy}
    G\left[c,\phi,\rho\right] = \int_V d\mathbf{x} \left( g_h(c) + \frac{1}{2}\kappa\left\lvert\nabla c\right\rvert^2
    - \frac{1}{2}\varepsilon(c) \left\lvert\nabla\phi\right\rvert^2 + \rho\phi \right)
\end{equation}
{ The first term is the homogeneous free energy of the system $g_h$ and the second term corresponds to the penalty gradient term~\cite{balluffi2005kinetics,van1979thermodynamic,cahn1959free,cahn1958free}, which is used to describe the phase separation of the material. The phenomenological parameter $\kappa$ controls the thickness of the interface between the formed phases and is linked to their interfacial tension. The third and fourth term describe the total electrostatic energy, where $\varepsilon(c)$ is the electric permittivity of the material as a function of the species concentration, $\rho$ is the mobile charge density.} For the homogeneous free energy term $g_h$ we chose the regular solution model~\cite{hildebrand1936solubility,zhou2019theory,bai2011suppression}
\begin{equation}
\label{eq:hom_free_energy_loc}
    g_h = \Omega c(1-c) + \frac{k_B T}{v}\left(c\ln c +(1-c)\ln(1-c)\right)
\end{equation}
where $\Omega$ controls the interaction between the species particles - positive (negative) $\Omega$ corresponds to attractive (repulsive)  interactions between the species. In the absence of particle-particle interactions, $v$ is the particle volume. Assuming local equilibrium, we define the chemical potential of the neutral species as the variational derivative of the Gibbs free energy~\cite{gelfand2000calculus,kondepudi2014modern}
\begin{equation}
\label{eq:chem_pot}
    \mu = \frac{\delta G}{\delta c} = \mu_h(c)-\kappa\nabla^2c - \frac{1}{2}\frac{\partial\varepsilon}{\partial c}\left\lvert \nabla\phi \right\rvert^2
\end{equation}
where $\mu_h=\partial g_h/\partial c$. Also, the chemical potential of the charged species that contribute to the charge density $\rho$ reads
\begin{equation}
\label{eq:chem_pot_charge}
    \mu_{\rho} = e\frac{\delta G}{\delta \rho} = e\phi
\end{equation}
Regarding the dielectric model, we assume a simple monotonically increasing phenomenological form 
\begin{equation}
\label{eq:perm_model}
    \varepsilon = \varepsilon_f\varepsilon_0e^{\gamma c}
\end{equation}
where $\gamma$ controls both the change and the curvature of the permittivity. This form has been previously used to model the electric-field induced phase separation of colloidal mixtures~\cite{Johnson2004,brosseau1994dielectric}.

\subsection{\label{subsec:transport}Transport}
We consider a phase separating material with both the permittivity and conductivity to be functions of species concentration. The equations to model the process are~\cite{deen1998analysis,sze2006physics,jackson2007classical}
\begin{subequations}
\label{eq:conservations}
\begin{equation}
    \frac{\partial c}{\partial t} = -\nabla\cdot \mathbf{j}
\end{equation}
\begin{equation}
    \frac{\partial \rho}{\partial t} = -\nabla\cdot \mathbf{j}_{\rho}
\end{equation}
\begin{equation}
    \frac{\delta G}{\delta \phi} = \rho - \nabla\cdot\mathbf{D} = 0
\end{equation}
\end{subequations}
where $\mathbf{j}$ and $\mathbf{j}_{\rho}$ are the species and electronic fluxes, respectively. { Based on the assumptions of local equilibrium~\cite{de2013non,kondepudi2014modern} and microscopic reversibility~\cite{onsager1931reciprocal_1,onsager1931reciprocal_2}, the constitutive relation for the fluxes and the dielectric displacement $\mathbf{D}$ are
\begin{subequations}\label{eq:fluxes}
\begin{equation}
    \mathbf{j} = -\frac{D(c)c}{k_BT}\nabla\frac{\delta G}{\delta c} = -\frac{D(c)c}{k_BT}\nabla \mu
\end{equation}
\begin{equation}
    \mathbf{j}_{\rho} = -\sigma(c) \nabla\frac{\delta G}{\delta \rho} = -\sigma(c) \nabla \phi
\end{equation}
\begin{equation}
    \mathbf{D} = \varepsilon(c)\mathbf{E} = -\varepsilon(c)\nabla \phi
\end{equation}
\end{subequations}}
Eqs.~\ref{eq:conservations}(a)-(b) are the conservative descent~\cite{adams2013large} of the free energy $G$ with respect to $c$ and $\rho$, while eq.~\ref{eq:conservations}(c) is an equilibrium point with respect to $\phi$. 

In eq.~\ref{eq:fluxes}(b), we consider the electric flux to be described by Ohm's law~\cite{sze2006physics} and we do not specify neither the mechanism of electric conduction nor the identity of charge carriers. This approach is similar to the model of leaky dielectrics that has been widely used in electrohydrodynamics~\cite{melcher1969electrohydrodynamics,saville1997electrohydrodynamics,melcher1981continuum}, where the conductivity is assumed to originate from dissolved ions. The model of eq.~\ref{eq:fluxes}(b) is known to be a special limit of Poisson-Nerst-Planck type models under large applied electric fields and low carrier densities~\cite{schnitzer2015taylor,bazant2015electrokinetics}. 

The functional form of the diffusivity $D\left(c\right)$ is taken that of a lattice gas model, where the transition state is influenced by excluded volume effects leading to $D\left(c\right)=D_0v\left(1-c\right)$~\cite{bazant2013theory}. The permittivity is given in eq.~\ref{eq:perm_model}. The conductivity is assumed to be linear in concentration $\sigma\left(c\right)=\sigma_0+\sigma_1 c$, where for $\sigma_1>0$ increasing charge carriers correspond to increasing conduction~\cite{sze2006physics}. While being simple, this specific form of $\sigma$ does not alter our conclusions. When phase separation occurs, we assume the two formed phases to have conductivities which differ by several orders of magnitude (metal-insulator/semi-conductor contact and vice versa~\cite{nadkarni2019modeling}). This large difference in combination with the phase separation due to electric fields leads to dielectric breakdown in our system. Finally, we close the description of the system using the following boundary conditions: 1) for the species number density we assume blocking electrodes, which translates into $\mathbf{j}\cdot\mathbf{n}=0$ along all domain boundaries; 2) for the potential $\phi$, we consider $\phi\left(0\right)=V$ and $\phi\left(L\right)=0$ to simulate the voltage drop across the cell, while on all the other boundaries we assume $\mathbf{n}\cdot\nabla\phi=0$; 3) we assume a contact angle of $\pi/2$ at any triple contact point between the formed phases and the boundaries of the system, $\mathbf{n}\cdot\nabla c=0$. In all cases, $\mathbf{n}$ is the outward normal vector.

\subsection{\label{subsec:dimnless}Characteristic Scales and Material Parameters}

The characteristic scales we consider herein are the following: (i) time $t_{ch}=L^2/D_{max}$, (ii) length $L_{ch}=L$, (iii) voltage $\phi_{ch}=k_BT/e$, (iv) conductivity $\sigma_{ch}=\sigma_{max}$, (v) charge density $\rho_{ch}=en_{max}N_A$, (vi) volumetric energy $k_BT/v$, where $N_A$ is the Avogadro number. Substituting these scales in the transport equations we arrive at the following dimensionless forms
\begin{subequations}
\label{eq:dimless_transport}
\begin{equation}
    \frac{\partial c}{\partial \tau} = \nabla\cdot\left[ \frac{D(c)}{D_{max}} \nabla{\mu}\right]
\end{equation}
\begin{equation}
    \frac{\partial q}{\partial \tau} = \nabla \cdot \left[\Phi \frac{\sigma(c)}{\sigma_{max}} \nabla \phi\right]
\end{equation}
\begin{equation}
    -\frac{1}{\lambda_D^2}\nabla\cdot\left[\varepsilon(c)\nabla\phi\right] = q
\end{equation}
\end{subequations}
where $\Phi=\frac{\sigma_{max}k_BT}{D_{max}n_{max}N_Ae^2}$, $\lambda_D^{-2}=\frac{\er k_BT}{n_{max}N_AL^2e^2}$. The dimensionless number $\Phi$ quantifies the ratio between the electronic and the species mobilities. An interesting limit is that of $\Phi\gg1$, where the electron/hole redistribution in the domain occurs much faster compared to species diffusion, and can be thought as the continuum equivalent of the Born-Oppenheimer approximation~\cite{kittel1976introduction}. In that case, eq.~\ref{eq:dimless_transport}(b) is always at quasi-equilibrium. This is not the case though when the material enters the insulating regime. Finally, the homogeneous free energy combined with the electric energy reads
\begin{equation}
\label{eq:hom_free_energy}
\begin{split}
    \frac{gv}{k_BT} =\frac{\Omega v}{k_BT}c\left(1-c\right) & + c\ln c +(1-c)\ln(1-c) \\ &
    - \frac{\er v k_BT}{2e^2 L^2}e^{\gamma c}\left\lvert\nabla\phi\right\rvert^2
\end{split}
\end{equation}

Before we dive further into analyzing the implications of the applied electric field on the thermodynamics stability of mixing and the dynamics of phase separation, it is instructive to consider a specific material for the parameters of our model: $\varepsilon_r$, $\gamma$, $\kappa$, $\Omega$, $v$, $n_{max}$, $\sigma_{max}$, $D_{max}$ and $L$. An interesting example with practical implications in Li-ion batteries and neuromorphic computing is Li$_{4+3x}$Ti$_5$O$_{12}$~\cite{vasileiadis2018toward,gonzalez2020lithium}. Based on the density and the molecular weight of Li$_{7}$Ti$_5$O$_{12}$, we know that $n_{max}\simeq 0.72\times 10^4$ mol/m$^3$ as well as $v \simeq 1$ nm$^3$. Additionally, previous phase field modeling~\cite{vasileiadis2018toward} has reported $N_A\Omega v\simeq8.612$ kJ/mol and $N_A\kappa v \simeq 8.61\times10^{-15}$ J m$^2$/mol. The electric permittivity is known to increasing as a function of average Li-ion fraction $x$. In particular, for $x\rightarrow0$, $\varepsilon\simeq1.5\varepsilon_0$, while for the fully lithiated state,  $x\rightarrow1$, the permittivity becomes $\varepsilon\simeq50\varepsilon_0$~\cite{liu2017first}. This behavior can be approximated by choosing $\varepsilon_r\simeq 1.5$ and $\gamma\simeq 3.5$. In terms of tracer diffusivity, NMR studies~\cite{wagemaker2009li,schmidt2015small,wilkening2007microscopic} measured $D_0\simeq 4\times10^{-16}$ m$^2/$s, while electrochemical measurements found that the conductivity changes from $\sigma(x\rightarrow0)\simeq 10^{-5}$ S$/$m (insulating) to $\sigma(x\rightarrow1)\simeq 10^{2}$S$/$m (metallic)~\cite{verde2016elucidating}. Finally, we consider a thin-film device with dimensions around $L\simeq100$ nm, which is kept at constant temperature $T=298$ K. 

The discussed material and system parameters result in $\Phi\simeq10^{7}$ and $\lambda_D^{-1}\simeq10^{-4}$. Given this value for $\Phi$, it is clear that when the material is metallic, we can assume eq.~\ref{eq:dimless_transport}(b) to be in quasi-equilibrium. The small value of $\lambda_D^{-1}\simeq10^{-4}$ corresponds to double layers on the scale of $10^{-1}$\si{\angstrom}, which is a reasonable value for a perfect metal. The RC timescale $\tau_C=\sqrt{\frac{e^2 n_{max}N_A\varepsilon_r\varepsilon_0L^2}{\sigma^2k_BT}}$ for charging the formed double layers after the electric field is applied~\cite{bazant2004diffuse} is of the order of $5\times10^{-10}\,\,s$ for the conductive phase and $10^{-2}\,\,s$ for the insulating one, values much lower than the diffusive timescale of the neutral species $\tau_D=L^2/D_{max}\sim 10^2\,\,s$. Due to numerical stability, however, when we solve eqs.~\ref{eq:dimless_transport} we use a re-scaled value for $\lambda_D^{-1}$ that matches the interface thickness. 

Finally, we discretize the set of eqs.~\ref{eq:dimless_transport} using finite elements~\cite{fraggedakis2017discretization}, and more specifically, we use continuous linear basis functions for approximating all unknowns. Additionally, we solve the system of equations in a monolithic fashion, while for the time integration a second order scheme is used (BDF2)~\cite{fraggedakis2015flow}. The non-linear system of equation is solved using Newton's method and for the inversion of the resulting linear system we use LU decomposition.

\section{\label{sec:demixing} Instability of a Homogeneous State}

\subsection{Thermodynamics Stability}

\begin{figure*}[!ht]
    \centering
    \hspace{0.08in}\includegraphics[width=1.\textwidth]{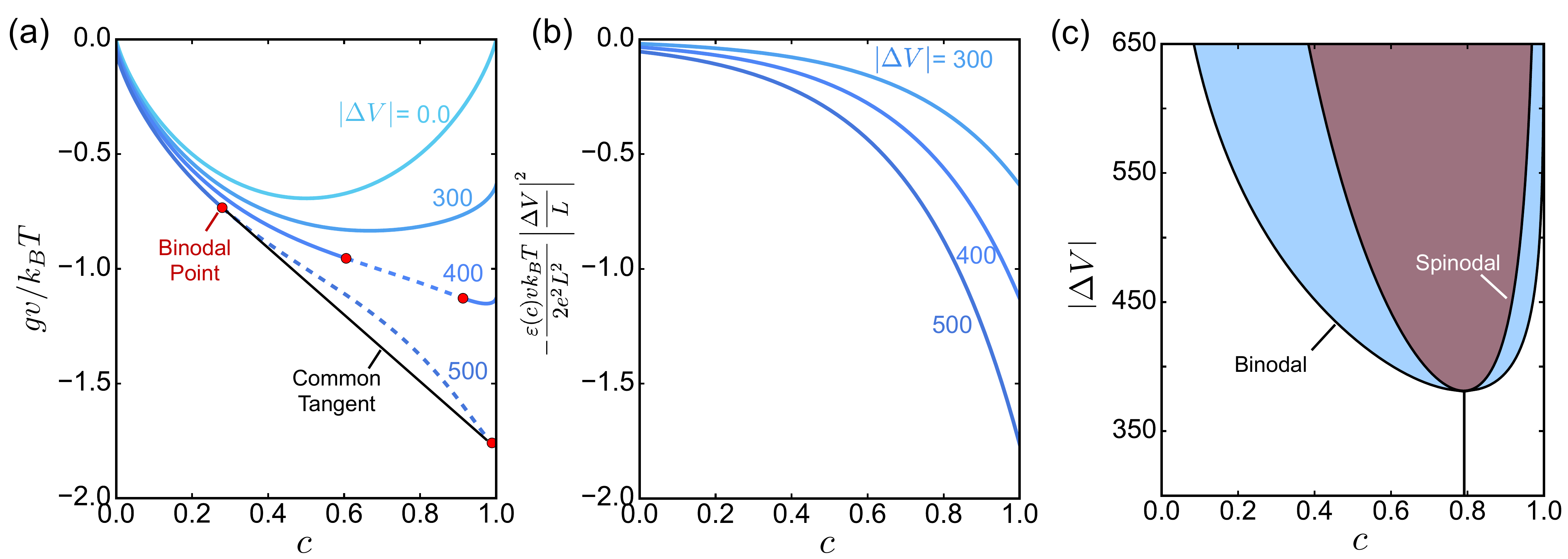}
    \caption{(a) Homogeneous free energy diagram as a function of the species concentration. Different curves correspond to different applied { dimensionless voltage drop $\Delta V$ (scaled to $k_BT/e$)} between the two electrodes. With increasing electric field, the free energy loses its convexity, indicating the formation of two phases with concentrations determined by the common tangent construction. { (b) The electrostatic contribution to the free energy for three different values of dimensionless voltage drop $\Delta V$.} (c) Thermodynamics phase diagram in terms of the species concentration and the applied voltage drop $\Delta V$. The light brown region corresponds to the miscibility gap, which changes with increasing voltage drop, and the blue region is the binodal region, where the solution is metastable. The present phase diagram is generated using the model of eq.~\ref{eq:hom_free_energy} with $\Omega = 0$.}
    \label{fig:free_energy_phase_diagram}
\end{figure*}

According to phase equilibria, $c_{s,1}$ and $c_{s,2}$ are the two spinodal points that indicate the change in the curvature of the homogeneous free energy. Both of these values are solutions of~\cite{kondepudi2014modern,chaikin1995principles}
\begin{equation}
    \frac{\partial^2 g_h}{\partial c^2} = \frac{\partial \mu_h}{\partial c} =0
\end{equation}
In the spinodal region, the homogeneous solution is unstable ($g_h''<0$) and tends to phase separate in two immiscible phases. The concentration in each phase is determined by the common tangent construction
\begin{equation}
\begin{split}
    \mu(c_{eq,1},\left\lvert\mathbf{E}\right\rvert=0)=&\mu(c_{eq,2},\left\lvert\mathbf{E}\right\rvert=0)=\\
    &\frac{g_h(c_{eq,1})-g_h(c_{eq,2})}{c_{eq,1}-c_{eq,2}}
\end{split}
\end{equation}

Here, we are interested in studying the electric-field induced phase separation. When the electric field is uniform across medium, $\mathbf{E}\simeq -\Delta V\,\mathbf{e}_x$ { ($\Delta V$ is the dimensionless voltage drop, scaled with the thermal voltage $k_BT/e$)}, we can see from eq.~\ref{eq:chem_pot} that the value of the chemical potential will change. The effect of the electric field on the thermodynamics stability of the homogeneous mixture, however, depends on the functional form of $\varepsilon$. For example, when $\varepsilon$ is a linear function of $c$, then the spinodal region is not affected. Therefore, for having electric field-induced phase separation we require that $\varepsilon''(c) \neq 0$. The equation for finding the spinodal points reads~\cite{tomlinson1972spinodal,thornburg1972electric,gonzalez2020lithium}
\begin{equation}
\label{eq:spinodals_e_field}
    \frac{\partial^2 g_h(c)}{\partial c^2} =\frac{1}{2}\frac{\partial^2 \varepsilon}{\partial c^2}\left\lvert\Delta V\right\rvert^2
\end{equation}

Fig.~\ref{fig:free_energy_phase_diagram}(a) shows the dimensionless free energy as a function of the species fraction $c$ for four different values of the applied electric field $\Df$. In this example we set $\Omega=0$, as we are interested to understand the implications of the electric field on the de-mixing of a homogeneous solution. For $\Df=0$, the energy is convex, and the mixture remains in the solid solution regime. With increasing $\Df$, however, the energy landscape becomes distorted, shifting the minimum energy toward $c\simeq0.8$. For $\Df=400$, the electric field is strong enough to change the convexity of $g$ (dashed region), making phase separation thermodynamically favorable. { At this point, the electrostatic energy becomes comparable to the entropy of mixing due to alignment of the microscopic dipoles in the medium, Figs.~\ref{fig:free_energy_phase_diagram}(a) \& (b).} As it will be shown in more detail in later sections, after phase separation is completed the system will consist of domains with low and high permittivity, respectively. { Further increase of the $\mathbf{E}$-field increases the distance between the binodal points (red dots) which leads to further increase in the dielectric mismatch between the phase separated domains, Fig.~\ref{fig:free_energy_phase_diagram}(a).}

These observations can be summarized in the phase diagram of Fig.~\ref{fig:free_energy_phase_diagram}(c), which is constructed in terms of the applied electric field $\Df$ and the system fractional concentration $c$. The binodal region, which is thermodynamically metastable, is shown with blue, while the spinodal region with light brown. It is clear that there exist a critical electric field (around $\Df_c\sim380$ for the parameters used herein) above which the convexity of the free energy changes and the homogeneous state of the material becomes thermodynamically unstable. The implications of this phase diagram on the dielectric breakdown of the material are discussed in Section~\ref{sec:filament}. 

\subsection{\label{sec:lin_stab} Linear Stability}

\begin{figure*}[!ht]
    \centering
    \hspace{0.08in}\includegraphics[width=0.95\textwidth]{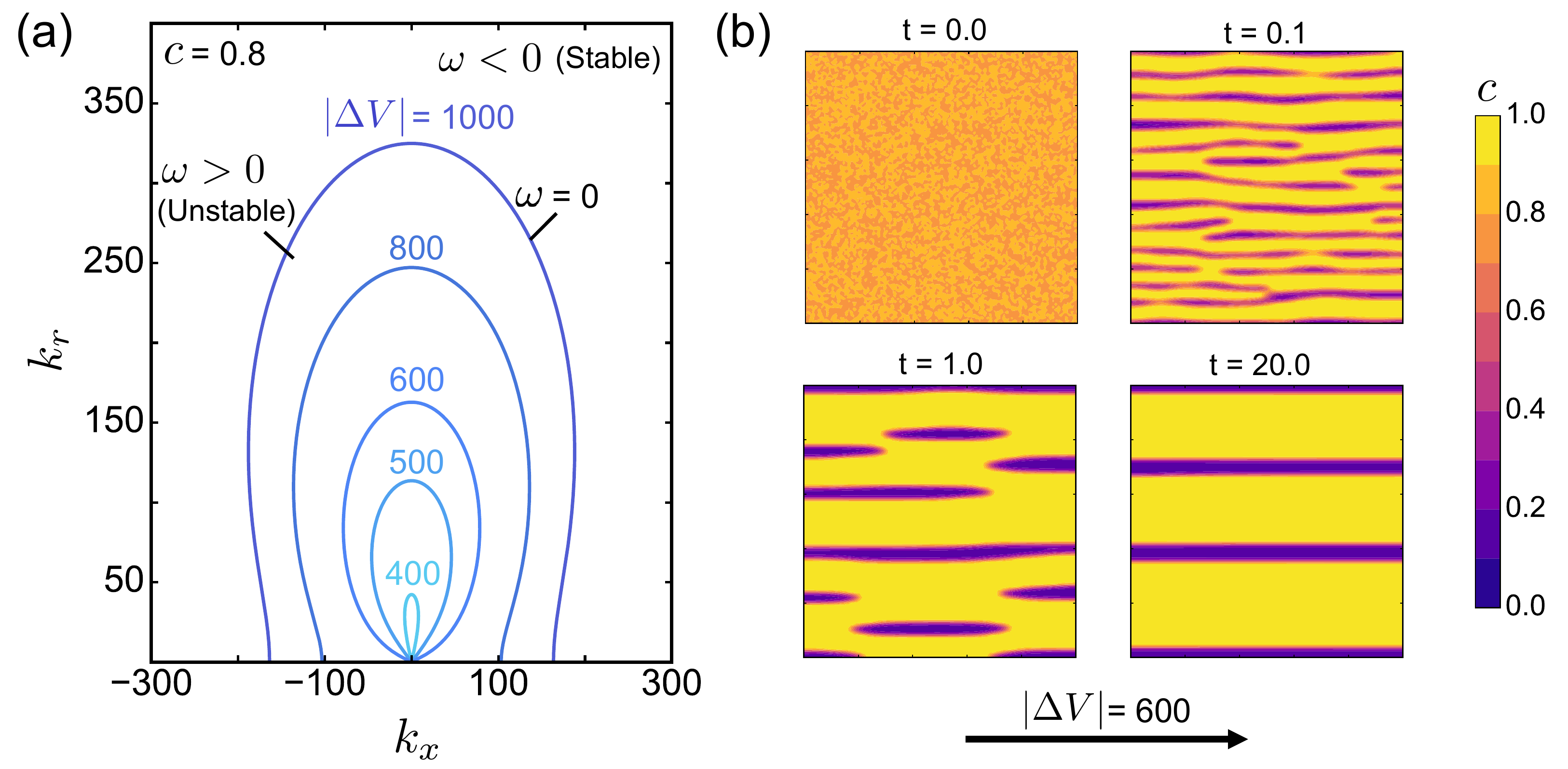}
    \caption{(a) Linear stability analysis of a homogeneous mixture with $c=0.8$ under the influence of an electric field $\mathbf{E}$. The contour lines correspond to the locus where the non-dimensional growth rate $\sd=0$ for different values of the wave numbers $k_x$ and $k_r=\sqrt{k_y^2+k_z^2}$. Different colored curves correspond to different values of the applied electric field. The thermodynamics model is described in eq.~\ref{eq:hom_free_energy} with $\Omega=0$. With increasing electric field, the homogeneous state becomes unstable for a larger set of $\left(k_x,k_r\right)$. (b) Temporal evolution of an initially unstable homogeneous state at $c=0.8$. A voltage drop of $\Df=600$ is applied. At first, the homogeneous state splits into multiple filaments with with concentrations $0.12$ and $0.998$. As time increases, the initially formed filaments accumulate into larger islands which at the end merge to form filaments that align with the applied electric field. 
    }
    \label{fig:linear_stability_dynamics}
\end{figure*}

The standard way to understand the dynamics during the onset of phase separation is through linear stability analysis. In particular, given the physical parameters of our model, we can identify the critical wavelength that can be induced by random fluctuations that lead to de-mixing of a homogeneous state. To do so, we assume an infinitesimal perturbation of the form $\delta \mathbf{y}=\bm{\delta} e^{i\mathbf{k}\cdot\mathbf{x}+\sd t}$, where $\mathbf{y}^T=\left(c,q,\phi\right)$, $\bm{\delta}$ is an infinitesimal vector, $\mathbf{k}^T=\left(k_x,k_y,k_z\right)$ is the wavenumber, and $\sd$ is the growth rate of the instability. The base state around which we linearize eqs.~\ref{eq:dimless_transport} is $\mathbf{y}_0^T=\left(c_0,0,-\Delta V\left(1-x\right)\right)$. The dispersion relation is found by solving the secular equation~\cite{leal2007advanced}
\begin{equation}
    \det\left\lvert\mathbf{J}-\sd\mathbf{e}_1\mathbf{e}_1\right\rvert=0
\end{equation}
where $\mathbf{J}$ is the Jacobian matrix defined as $\mathbf{J}=\delta_\mathbf{y} f$ - the components of $\mathbf{J}$ are given in the appendix - and $\mathbf{e}_1$ corresponds to the unit vector along the `concentration' axis. Assuming $\Phi\gg1$, the growth rate becomes
\begin{equation}
\label{eq:dispersion}
\begin{split}
    \sd = & - \varepsilon\Df^2\left(\partial_c\ln\sigma\right) k_x^2 -D\left(k_x^2+k_r^2\right)\times\\
    & \left[
    -\left(1/2\right)\Df^2\partial^2_c\varepsilon
    +\partial^2_c g_h
    +\kappa\left(k_x^2+k_r^2\right)
    \right]
\end{split}
\end{equation}
where $k_r^2=k_y^2+k_z^2$, and $D$, $\sigma$ and $\kappa$ are in their dimensionless form. From eq.~\ref{eq:dispersion}, it is clear that the direction of the $\mathbf{E}$-field affects critical wavelength in the x-direction. We can further analyze the result by determining the set of $\left(k_x,k_r\right)$ that maximize $\sd$. Solving $\partial \sd/\partial \mathbf{k}=0$, we find
\begin{subequations}
\label{eq:k_max}
\begin{equation}
    \left(k_x,k_r\right)=\left(0,\sqrt{\frac{Q}{4\kappa}}\right)
\end{equation}
\begin{equation}
    \left(k_x,k_r\right)=\left(\pm \sqrt{\frac{DQ-2\varepsilon\Df^2\partial_c\ln\sigma}{4\kappa D}},0\right)
\end{equation}
\end{subequations}
where $Q = \Df^2\partial_c^2\varepsilon-2\partial_c^2g_h$. From the first set of solution, it is clear that for $k_r$ to be physical it has to be positive. This is true for $\Df^2\partial_c^2\varepsilon-2\partial_c^2g_h>0$, which is equivalent to the thermodynamics stability condition we discuss in Sec.~\ref{sec:demixing}A. On the contrary, the second locus of solutions is physical for $D\left(\Df^2\partial_c^2\varepsilon-2\partial_c^2g_h\right)>2\varepsilon\Df^2\partial_c\ln\sigma$, which shows that conductivity variations can suppress phase separation in the direction of electric field when $\partial_c\ln\sigma>0$.

Fig.~\ref{fig:linear_stability_dynamics}(a) demonstrates the stability diagram for a mixture with average concentration $c=0.8$ in terms of the wavenumber set $\left(k_x,k_r\right)$. The lines denote the isocontour $\sd=0$ for different applied voltages across the domain. Inside the formed envelopes lies the region where $\sd>0$, which corresponds to the long-wave modes for which de-mixing occurs, while short-waves are damped by the action of surface tension. As shown by eq.~\ref{eq:k_max}(b), when $k_r=0$ there is a critical applied voltage - $\Df\lesssim700$ below which perturbation in the $x$ direction are suppressed.

\subsection{\label{sec:phase_sep_dynamics} Phase Separation Dynamics}

In order to test the predictions of the theory and understand the time evolution of de-mixing, we perform two-dimensional simulations of an initially homogeneous mixture with concentration $c=0.8$. As a representative example, we consider the case where the value of the applied electric field across the domain is $\Df=600$. According to the phase diagram of fig.~\ref{fig:free_energy_phase_diagram}(c), we expect the two formed phases to have concentrations $c_{b,1}\simeq0.12$ and $c_{b,2}\simeq0.998$, respectively, where $c_{b}$ is the binodal point concentration. 

Fig.~\ref{fig:linear_stability_dynamics}(b) shows the temporal evolution of the concentration field after the electric field is applied. The light yellow color represents the rich phase, while the dark purple the low concentration one. At time $t=0$, the homogeneous profile is perturbed with zero-mean white noise. After some time, these perturbations grow exponentially in time as predicted by the linear stability analysis of Sec.~\ref{sec:lin_stab}. The exponential evolution of the instability stops right after the two phases begin to form, i.e. for $t=0.1$. At this moment, filament-like patterns, which consist of the poor and rich phase, span the entire domain. It is noticeable that the formed filaments align with the direction of the applied electric field, an observation that connects with the phenomenon of dielectric breakdown, which is discussed in more detail in the next section (Sec.~\ref{sec:filament}). At later times, $t=1.0$, the initially formed filaments with the lowest concentration break into smaller islands. Due to the existence of multiple interfaces this state is not energetically favorable. As a result, the system undergoes further coarsening (Ostwald ripening~\cite{voorhees1985theory}) making the islands to merge into larger filamentary domains, leading to the non-equilibrium steady state shown for $t=20.0$. At this time, three large domains that consist of the high concentration phase are formed, while the low concentration ones occupy smaller fraction of the total volume. This is in agreement with the theoretically predicted phase diagram, fig.~\ref{fig:free_energy_phase_diagram}(c), where the lever rule predicts that the phase with the lowest concentration occupies $\sim22 \%$ of the total system. { Finally, when the electric field is removed, the homogeneous free energy becomes convex again, leading to mixing of the two formed phases. Therefore, the recovery of a solid solution after the electric field bias is removed corresponds to volatile behavior~\cite{ielmini2018memory}.}

\section{\label{sec:filament} Dielectric Breakdown due to Filament Formation }

\begin{figure*}[!ht]
    \centering
    \hspace{0.08in}\includegraphics[width=0.9\textwidth]{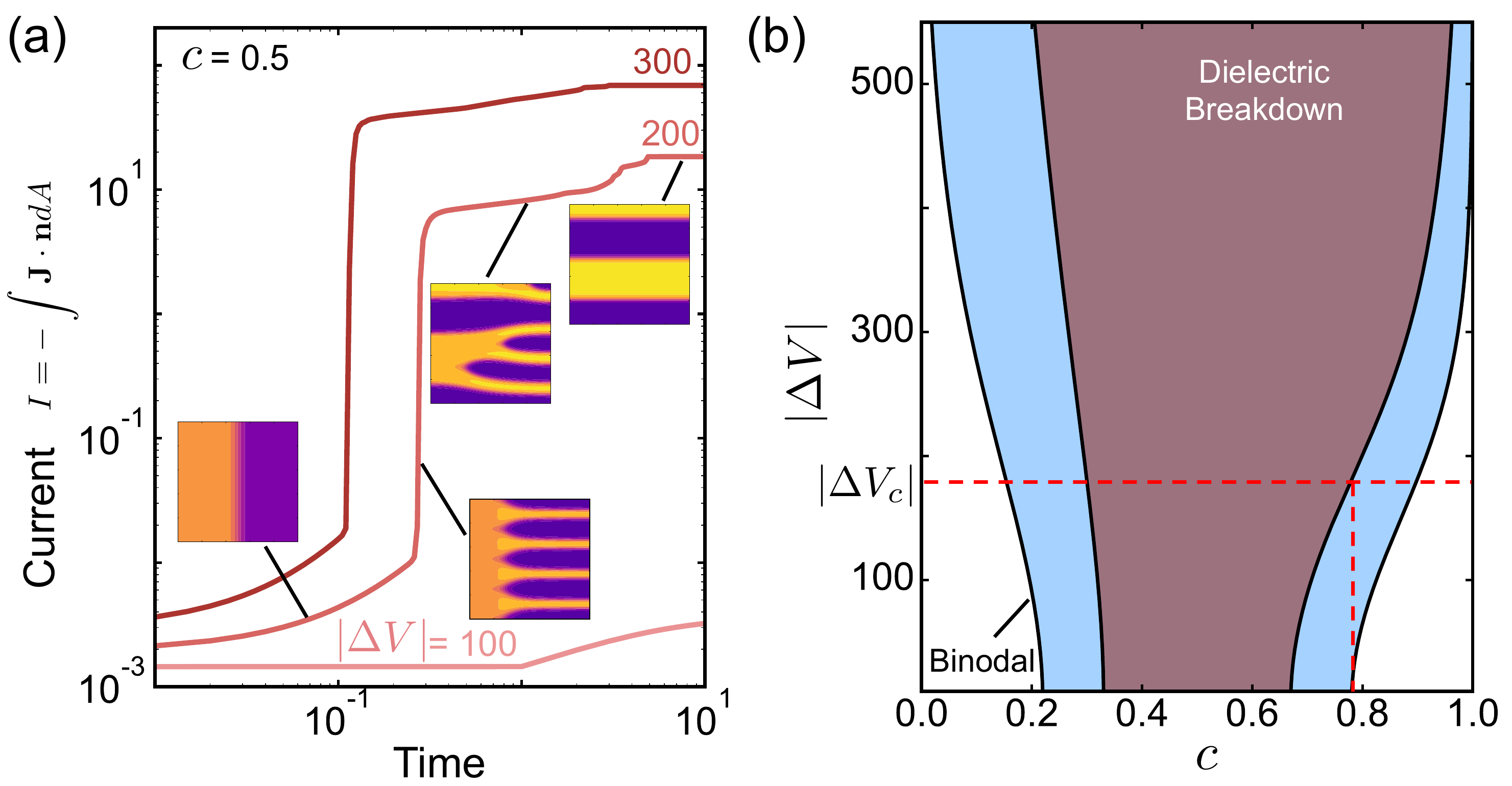}
    \caption{(a) Temporal evolution of macroscopic current defined as $I=-\int\sigma\mathbf{E}\cdot\mathbf{n}dA$, in response to an applied voltage $\Df$ (scaled to $k_BT/e$), indicated by different colors. The system is initially at a phase separated state as shown in the inset for the earliest time instance. The average concentration is $c=0.5$. At low voltage drop, e.g. $\Df=100$, the initial morphology remains intact. For $\Df$ above the critical value, any interface perpendicular to the applied $\mathbf{E}$-field is unstable. This leads to the formation of filaments that connect the two electrodes and increase the electrical conductivity of the device by orders of magnitude, causing effectively dielectric breakdown. (b) Thermodynamics phase diagram in terms of the species concentration and the applied voltage drop $\Delta V$. The light brown region corresponds to the miscibility gap, which changes with increasing voltage drop, and the blue region is the binodal region, where the solution is metastable. The critical voltage drop $\Delta V_c$ is defined as the value of $\Df$ that shifts one of the binodal points at zero bias into the spinodal region. For both the simulations and the phase diagram, $N_A\Omega v = 5.601$ kJ/mol, while the electric conductivity is described by $\sigma=\sigma_0e^{\sigma_1 c}$, where $\sigma_1=10^{-7}$ and $\sigma_1=16.11$.}
    \label{fig:dielectric_breakdown}
\end{figure*}

The de-mixing of an initially solid-solution system demonstrates the basic principle of the electric-field induced phase separation. Although studying these materials is informative, many materials of practical relevance phase separate at room temperature, even in the absence of $\mathbf{E}$-field. Such examples are Li$_x$CoO$_2$, Li$_{4+3x}$Ti$_5$O$_{12}$, and Li$_x$TiO$_2$ where during phase separation undergo metal-to-insulator (and vice versa) transition. Here, we focus on the case of an initially phase separated material and show that, by applying electric field, it is possible to control the orientation of the phase boundaries and, consequently, the current-voltage response of the material.

For the ease of computations and without altering the final conclusions, we choose a system with $N_A\Omega v\simeq 5.601$ kJ/mol, and electrical conductivity of the form $\sigma=\sigma_0e^{\sigma_1 c}$, where $\sigma_1=10^{-7}$ and $\sigma_1=16.11$. All the other properties are the same as discussed in Sec.~\ref{subsec:dimnless}. The reason for changing the functional form of $\sigma$ is to replicate the abrupt change in the electrical conductivity during the insulator-to-metal transition that take place in materials like Li$_{4+3x}$Ti$_5$O$_{12}$~\cite{verde2016elucidating,wagemaker2009li}.

Fig.~\ref{fig:dielectric_breakdown}(a) demonstrates the temporal evolution of the resulting current for three different applied $\Delta V$. The current is defined as the surface integral of the electric current density across one of the electrodes, $I=-\int{\mathbf{J}\cdot\mathbf{n}}dA$. For all the applied voltages, the initial phase morphology corresponds to the earliest time instant shown in the inset of Fig.~\ref{fig:dielectric_breakdown}(a), while the average concentration is set to $c=0.5$. 

For $\Delta V=100$, the resulting current is always of the order of $10^{-3}$, which can be understood in terms of an equivalent circuit. Given the functional form for $\sigma$, we know that one of the phases is insulating. Also, the phase morphology does not change after the voltage is applied. Therefore, the equivalent circuit consists of two resistances in series, one of which corresponds to an insulator with resistance $R=\Delta V/I \sim O\left(10^5\right)$. 

For larger applied voltages, $\Delta V\gtrsim 200$, the temporal evolution of the current is qualitatively different. More specifically, we focus on the phase evolution for $\Delta V=200$. It is apparent from the inset images that the applied electric field across the domain is able to change the morphology completely. At early times, $t_c<0.27$, the electric field forces the binodal concentration to change, as shown in the phase diagram of fig.~\ref{fig:dielectric_breakdown}(b). Due to this change, the system is perturbed and the interface between the two phases becomes unstable forming tips in the direction of the electric field. At around $t_c\simeq 0.27$, the instability grows abruptly leading to the formation of highly conductive filaments. By the time these filaments `touch' the second electrode, the electric current increases by three orders of magnitude, from $I\simeq 0.01$ to $I\simeq 7$. After this critical event, a dendrite-like structure is formed, $t\simeq1$, which evolves in four filaments for $t\gtrsim5$ - two consisting of the highly conductive phase and two of the insulating one. The insulating domains within the dendrite-like structure demonstrate a specific angle that can be related to the Taylor cone behavior~\cite{taylor1964disintegration}. The final configuration yields a steady state current around $I\simeq19$, which exceeds by almost four orders of magnitude the value obtained when $\Delta V=100$ is applied. Therefore, we conclude that the formation of highly electric conductive filaments after a critical voltage is applied leads to the phenomenon of dielectric breakdown.

The question that arises, though, is why voltages below $\Delta V\simeq200$ do not alter the phase morphology into filamentary ones. The answer becomes clear when analyzing the phase diagram of fig.~\ref{fig:dielectric_breakdown}(b). When no $\mathbf{E}$-field is applied, the system is separated in two immiscible phases with concentration $c_{b,1}\simeq0.21$ and $c_{b,2}\simeq0.78$, respectively. When a voltage drop of $\Delta V=100$ is applied, these values lie the binodal region, which is known to be metastable. As a result, the species are redistributed between the two phases and a new steady state is reached without altering the morphology of the existing interface. However, when $\Delta V=200$ is applied, the initially stable state becomes thermodynamically unstable, as the concentration of the rich phase lies inside the spinodal region - the spinodal points are $c_{s,1}\simeq0.3$ and $c_{s,2}\simeq0.8$. Hence, any infinitesimal perturbation tends to destabilize the system from its initial state, which, as analyzed in fig.~\ref{fig:dielectric_breakdown}(a) leads to the formation of highly conductive filaments that align with the direction of the applied electric field.

\section{\label{sec:disc}Discussion}

\begin{figure}[!ht]
    \centering
    \hspace{0.08in}\includegraphics[width=0.4\textwidth]{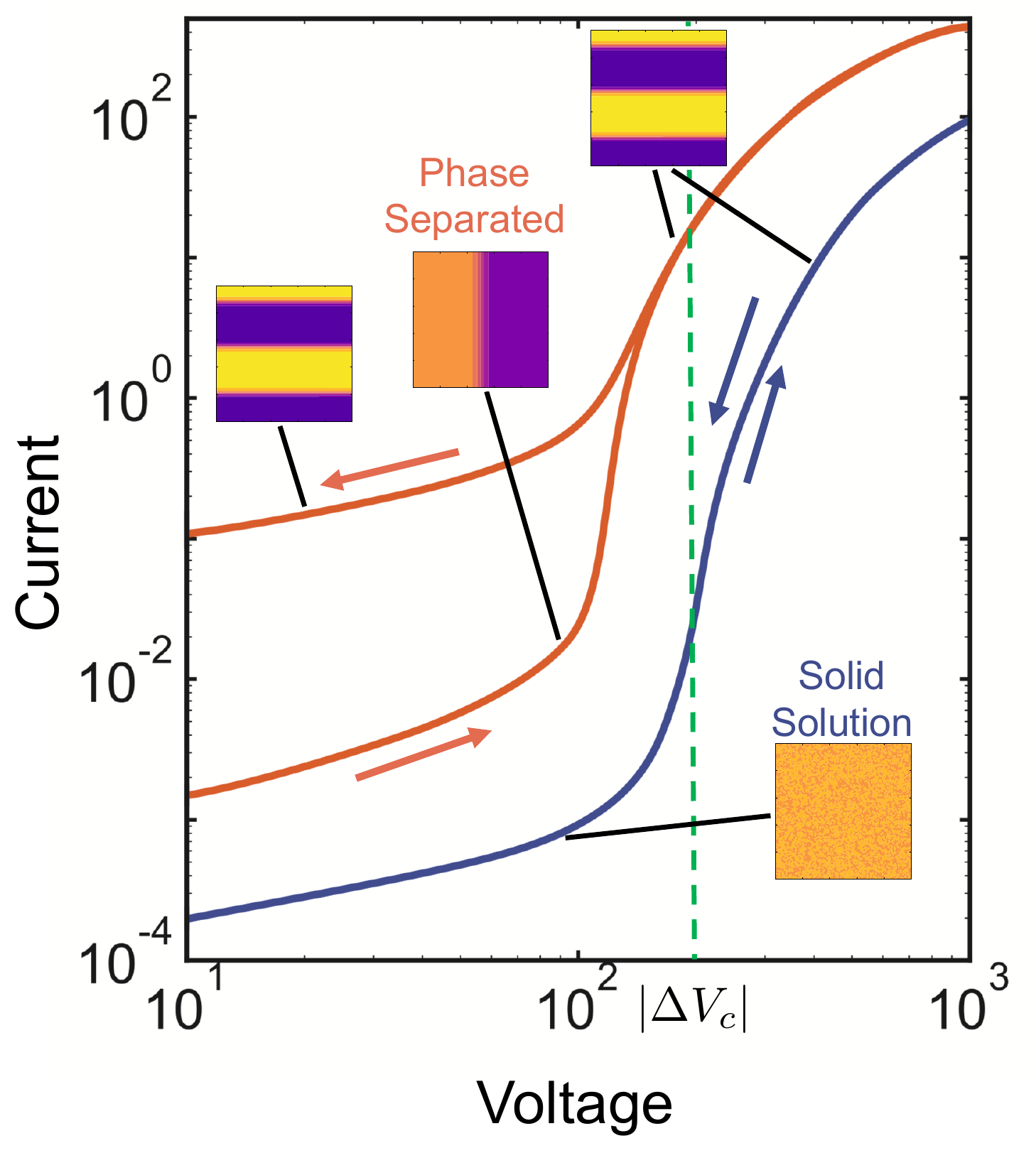}
    \caption{Schematic of the device-level current-voltage response for a mixture with electric permittivity that depends on species concentration $\varepsilon\left(c\right)$. Both curves correspond to the parameters used in Secs.~\ref{sec:demixing}-\ref{sec:filament}.
    The blue curve corresponds to a system initially prepared as a solid solution mixture, while the orange line is that for a phase separated system. The arrows indicate the direction of the voltage sweep. In both cases, there is a critical applied voltage above which filaments are formed between the two electrodes and the resistance of the device drops by orders of magnitude. This phenomenon corresponds to resistive switching, which is the key operation for the operation of memristors.}
    \label{fig:device_level}
\end{figure}

Our findings on dielectric breakdown, as a result of filament formation due to the applied electric field, are relevant to resistive switching and memristor devices~\cite{Zhou2015,Bisri2017,Lee2015,Wang2020,Sun2019,Green1928,ielmini2018memory,waser2019introduction}. One of the desired properties of memristors is the ability to change the electrical conductivity of the device by orders of magnitude under an external stimulus, e.g. voltage or temperature. Here, we show that through de-mixing of a solid solution mixture or change in the phase morphology of an already phase separated material, we can tune the device-level resistance by orders of magnitude, fig.~\ref{fig:dielectric_breakdown}(a). This idea has been recently demonstrated experimentally in Li$_{4+3x}$Ti$_5$O$_{12}$, for $x\rightarrow0$ and $x\rightarrow1$, a material proposed as a potential candidate for memristive devices~\cite{gonzalez2020lithium}. It is known that Li$_{4+3x}$Ti$_5$O$_{12}$ undergoes an insulator-to-metal transition~\cite{verde2016elucidating}, resulting in a change in the electric resistance of the memristive device by orders of magnitude. Here, by using the properties of LTO, we show that a dimensionless voltage of $\sim O\left(10^2\right)$ is required to induce de-mixing, and therefore change in the conductivity of the device. This value corresponds to approximately $3-4$ V, which is on the same order of magnitude with the experimentally observed value for dielectric breakdown. 

Fig.~\ref{fig:device_level} shows a schematic on the device current-voltage response related to the phenomena presented in Secs.~\ref{sec:demixing}-\ref{sec:filament}. When the system is initially prepared as solid solution, de-mixing occurs at a critical voltage $\Delta V_c$, figs.~\ref{fig:free_energy_phase_diagram}(b) \&~\ref{fig:dielectric_breakdown}(b). Two scenarios are possible: i) the mixture shows solid solution behavior for all values of $c$, ii) the mixture is phase separating for a wide range of concentrations. In the former situation, when the applied voltage is close to critical value the two phases have very similar concentration, and thus, similar conductivities. Therefore, for resistive switching applications a voltage much larger than the critical value needs to be applied in order to establish phases with very large mismatch in their conductivity. For the latter case, when the critical voltage is applied and de-mixing occurs, the current changes abruptly due to the large difference on the electric resistance between the formed phases, fig.~\ref{fig:dielectric_breakdown}(a). Although dielectric breakdown is necessary for resistive switching, memristors have the additional requirement of non-volatile operation~\cite{ielmini2018memory}. When the electric field bias is lifted, the de-mixed mixture is going to return to its initial homogeneous state under diffusive timescales. In particular, for a device with $L=100$ nm~\cite{gonzalez2020lithium}, and a material with maximum diffusion of $\sim 10^{-16}$ m$^2$/s~\cite{vasileiadis2018toward}, the system relaxes back to equilibrium within $\sim 100$ seconds. This timescale is very short for any application that requires non-volatile operation, such as neuromorphic computing.
 
{ On the contrary, mixtures that are thermodynamically unstable under no applied electric field are non-volatile due to the persistence of the filaments after we stop applying a voltage bias.} A schematic of the representative current-voltage curve is shown with the orange line in fig.~\ref{fig:device_level}. At first, the system is prepared in a state where the interface between the two phases is aligned perpendicular to the direction of the electric field. As discussed earlier, this configuration corresponds to a circuit with two resistors in series, where the one with the lowest conductivity governs the $I-V$ response of the device. At large enough voltages, though, the phase separated mixture becomes thermodynamically unstable and conductive filaments are formed that connect the two electrodes. At this point, the macroscopic resistance of the device drops by orders of magnitude causing an effective dielectric breakdown - from a high resistance state (HRS) to a low resistance one (LRS)~\cite{ielmini2018memory,waser2019introduction}. After filament formation, decreasing the voltage drop does not affect the phase morphology. Hence, the persistence of the filamentary state, even after electric fields are not active, demonstrate the potential application of such systems in neuromorphic computing.

Even though our model is able to predict the correct order of magnitude for the critical voltage that causes dielectric breakdown in LTO, it is greatly simplified. First, the functional forms for both the electric permittivity and conductivity are purely empirical and do not describe an actual material. Therefore, a more complete theory for both $\varepsilon$ and $\sigma$, which takes into account information from first-principles calculations and/or experiments, is needed. 

A more complete picture would identify the charged species, e.g. bound and/or free electrons or ions, that can contribute to the electric conductivity of the medium. For example, in the case of quasi-particles such as polaron-ion pairs, if the applied electric field is strong enough, e.g. near electrode/bulk interfaces~\cite{luntz2015interfacial} or phase boundaries~\cite{Tsori2004}, the pairs can split into its components, i.e. localized electrons and ions. Each of the newly generated species can have its own conductivity~\cite{maxisch2006ab,malik2013critical} where its diffusive/conductive motion would be described by the corresponding conservation law.

Another effect we have neglected, which plays an important role when the system is at its LRS, is Joule heating~\cite{kim2011nanofilamentary,ielmini2011modeling,strachan2011switching,pickett2013scalable,Wang2019a,Wang2019b}. For nanometer scale phase change memristors, Joule heating is known to control resistive switching. Hence, the change of the local temperature due to dissipation phenomena, such as electric conduction, is expected to affect the thermodynamics and, consequently, the phase separation dynamics after the electric field is applied.

Most memristive devices are solid state in nature~\cite{kim2011nanofilamentary,ielmini2018memory,waser2019introduction}. { Thus, it is expected that elastic and/or inelastic deformation, as well as the existence of grain boundaries and dislocations to influence the dynamics of conductive filament formation.} Additionally, phase-separating intercalation materials are known to exhibit misfit strains which affect the morphology of the formed interfaces~\cite{cogswell2012coherency,nadkarni2018interplay}. Therefore, there will be a competition between the interface orientation defined by the minimum elastic energy state and the one induced by imposed electric field. 

Finally, in our model we assumed a closed system where species concentration does not change. However, in ion intercalation materials one can change the total number of ions. This is known to have a large impact on the phase morphology~\cite{nadkarni2018interplay,bai2011suppression,bai2011suppression} as well as on the electronic conductivity of the material~\cite{nadkarni2019modeling}, e.g. metal-to-insulator transition. Therefore, it would be interesting to explore the effects of species insertion/extraction on the phase morphologies at the same time electric fields are applied. All these phenomena have to be examined in greater detail for establishing qualitative design principles for memristive devices that are based on the phenomenon of the electric-field induced phase separation.

\section{\label{sec:conc}Summary}

In summary, we showed that when electric field is applied in a material with concentration-dependent permittivity and electric conductivity, phase separation occurs and dielectric breakdown is observed. Through thermodynamics stability analysis we derived phase diagrams in terms of the species concentration and the applied voltage drop between the operating electrodes, and we demonstrated that one can de-mix a solid solution mixture. Additionally, by performing simulations we predicted that once the system is thermodynamically unstable, filament-like structures are formed. These structures percolate across the domain and are responsible for the dielectric breakdown by allowing electrons to conduct through the metallic phase. Furthermore, we demonstrated the predictions of the theory to be in agreement with recent experiments on Li$_{4+3x}$Ti$_{5}$O$_{12}$. Finally, we discussed the implications of our results on resistive switching, which can be useful in applications like neuromorphic computing. In particular, we showed that phase separating materials can exhibit the desired non-volatile behavior while solid solution materials do not, as they relax back to their equilibrium state after the electric field is turned off. 

\section*{Acknowledgment}
The authors would like to thank Tao Gao, Neel Nadkarni, Juan Carlos Gonzalez-Rosillo, Moran Balaish, and Jennifer L. M. Rupp for insightful discussions.

\section*{\label{sec:appendix}Appendix}

\subsection*{\label{sec:jacobian}Jacobian Matrix}

The components of the Jacobian matrix discussed in Sec.~\ref{sec:lin_stab} are presented here. More specifically,
\begin{equation}
\mathbf{J} =     
\begin{bmatrix}
J_{1,1} & 0 & J_{1,3}\\
J_{2,1} & 0 & J_{2,3}\\
J_{3,1} & J_{3,2} & J_{3,3}
\end{bmatrix}
\end{equation}
where
\begin{subequations}
\begin{equation}
    J_{1,1} = -k^2Dc\left(\partial_c^2g_h+\kappa k^2-\frac{1}{2}\partial_c^2\varepsilon\left\lvert\nabla\phi\right\rvert^2\right)
\end{equation}
\begin{equation}
    J_{1,3} = i\partial_c\varepsilon k^2\mathbf{k}\cdot\nabla\phi
\end{equation}
\begin{equation}
    J_{2,1} = i\partial_c\sigma\mathbf{k}\cdot\nabla\phi
\end{equation}
\begin{equation}
    J_{2,3} = -\sigma k^2
\end{equation}
\begin{equation}
    J_{3,1} = -i\partial_c\varepsilon\mathbf{k}\cdot\nabla\phi
\end{equation}
\begin{equation}
    J_{3,2} = \lambda_D^2
\end{equation}
\begin{equation}
    J_{3,3} = \varepsilon k^2
\end{equation}
\end{subequations}

\bibliography{literature}% Produces the bibliography via BibTeX.

\begin{thebibliography}{98}%
\makeatletter
\providecommand \@ifxundefined [1]{%
 \@ifx{#1\undefined}
}%
\providecommand \@ifnum [1]{%
 \ifnum #1\expandafter \@firstoftwo
 \else \expandafter \@secondoftwo
 \fi
}%
\providecommand \@ifx [1]{%
 \ifx #1\expandafter \@firstoftwo
 \else \expandafter \@secondoftwo
 \fi
}%
\providecommand \natexlab [1]{#1}%
\providecommand \enquote  [1]{``#1''}%
\providecommand \bibnamefont  [1]{#1}%
\providecommand \bibfnamefont [1]{#1}%
\providecommand \citenamefont [1]{#1}%
\providecommand \href@noop [0]{\@secondoftwo}%
\providecommand \href [0]{\begingroup \@sanitize@url \@href}%
\providecommand \@href[1]{\@@startlink{#1}\@@href}%
\providecommand \@@href[1]{\endgroup#1\@@endlink}%
\providecommand \@sanitize@url [0]{\catcode `\\12\catcode `\$12\catcode
  `\&12\catcode `\#12\catcode `\^12\catcode `\_12\catcode `\%12\relax}%
\providecommand \@@startlink[1]{}%
\providecommand \@@endlink[0]{}%
\providecommand \url  [0]{\begingroup\@sanitize@url \@url }%
\providecommand \@url [1]{\endgroup\@href {#1}{\urlprefix }}%
\providecommand \urlprefix  [0]{URL }%
\providecommand \Eprint [0]{\href }%
\providecommand \doibase [0]{http://dx.doi.org/}%
\providecommand \selectlanguage [0]{\@gobble}%
\providecommand \bibinfo  [0]{\@secondoftwo}%
\providecommand \bibfield  [0]{\@secondoftwo}%
\providecommand \translation [1]{[#1]}%
\providecommand \BibitemOpen [0]{}%
\providecommand \bibitemStop [0]{}%
\providecommand \bibitemNoStop [0]{.\EOS\space}%
\providecommand \EOS [0]{\spacefactor3000\relax}%
\providecommand \BibitemShut  [1]{\csname bibitem#1\endcsname}%
\let\auto@bib@innerbib\@empty
%</preamble>
\bibitem [{\citenamefont {Lim}\ \emph {et~al.}(2016)\citenamefont {Lim},
  \citenamefont {Li}, \citenamefont {Alsem}, \citenamefont {So}, \citenamefont
  {Lee}, \citenamefont {Bai}, \citenamefont {Cogswell}, \citenamefont {Liu},
  \citenamefont {Jin}, \citenamefont {Yu}, \citenamefont {Salmon},
  \citenamefont {Shapiro}, \citenamefont {Bazant}, \citenamefont {Tyliszczak},\
  and\ \citenamefont {Chueh}}]{lim2016origin}%
  \BibitemOpen
  \bibfield  {author} {\bibinfo {author} {\bibfnamefont {J.}~\bibnamefont
  {Lim}}, \bibinfo {author} {\bibfnamefont {Y.}~\bibnamefont {Li}}, \bibinfo
  {author} {\bibfnamefont {D.~H.}\ \bibnamefont {Alsem}}, \bibinfo {author}
  {\bibfnamefont {H.}~\bibnamefont {So}}, \bibinfo {author} {\bibfnamefont
  {S.~C.}\ \bibnamefont {Lee}}, \bibinfo {author} {\bibfnamefont
  {P.}~\bibnamefont {Bai}}, \bibinfo {author} {\bibfnamefont {D.~A.}\
  \bibnamefont {Cogswell}}, \bibinfo {author} {\bibfnamefont {X.}~\bibnamefont
  {Liu}}, \bibinfo {author} {\bibfnamefont {N.}~\bibnamefont {Jin}}, \bibinfo
  {author} {\bibfnamefont {Y.-s.}\ \bibnamefont {Yu}}, \bibinfo {author}
  {\bibfnamefont {N.~J.}\ \bibnamefont {Salmon}}, \bibinfo {author}
  {\bibfnamefont {D.~A.}\ \bibnamefont {Shapiro}}, \bibinfo {author}
  {\bibfnamefont {M.~Z.}\ \bibnamefont {Bazant}}, \bibinfo {author}
  {\bibfnamefont {T.}~\bibnamefont {Tyliszczak}}, \ and\ \bibinfo {author}
  {\bibfnamefont {W.~C.}\ \bibnamefont {Chueh}},\ }\href@noop {} {\bibfield
  {journal} {\bibinfo  {journal} {Science}\ }\textbf {\bibinfo {volume}
  {353}},\ \bibinfo {pages} {566} (\bibinfo {year} {2016})}\BibitemShut
  {NoStop}%
\bibitem [{\citenamefont {Ielmini}\ and\ \citenamefont
  {Wong}(2018)}]{ielmini2018memory}%
  \BibitemOpen
  \bibfield  {author} {\bibinfo {author} {\bibfnamefont {D.}~\bibnamefont
  {Ielmini}}\ and\ \bibinfo {author} {\bibfnamefont {H.-S.~P.}\ \bibnamefont
  {Wong}},\ }\href@noop {} {\bibfield  {journal} {\bibinfo  {journal} {Nature
  Electronics}\ }\textbf {\bibinfo {volume} {1}},\ \bibinfo {pages} {333}
  (\bibinfo {year} {2018})}\BibitemShut {NoStop}%
\bibitem [{\citenamefont {Fuller}\ \emph {et~al.}(2017)\citenamefont {Fuller},
  \citenamefont {Gabaly}, \citenamefont {L{\'e}onard}, \citenamefont {Agarwal},
  \citenamefont {Plimpton}, \citenamefont {Jacobs-Gedrim}, \citenamefont
  {James}, \citenamefont {Marinella},\ and\ \citenamefont
  {Talin}}]{fuller2017li}%
  \BibitemOpen
  \bibfield  {author} {\bibinfo {author} {\bibfnamefont {E.~J.}\ \bibnamefont
  {Fuller}}, \bibinfo {author} {\bibfnamefont {F.~E.}\ \bibnamefont {Gabaly}},
  \bibinfo {author} {\bibfnamefont {F.}~\bibnamefont {L{\'e}onard}}, \bibinfo
  {author} {\bibfnamefont {S.}~\bibnamefont {Agarwal}}, \bibinfo {author}
  {\bibfnamefont {S.~J.}\ \bibnamefont {Plimpton}}, \bibinfo {author}
  {\bibfnamefont {R.~B.}\ \bibnamefont {Jacobs-Gedrim}}, \bibinfo {author}
  {\bibfnamefont {C.~D.}\ \bibnamefont {James}}, \bibinfo {author}
  {\bibfnamefont {M.~J.}\ \bibnamefont {Marinella}}, \ and\ \bibinfo {author}
  {\bibfnamefont {A.~A.}\ \bibnamefont {Talin}},\ }\href@noop {} {\bibfield
  {journal} {\bibinfo  {journal} {Advanced Materials}\ }\textbf {\bibinfo
  {volume} {29}},\ \bibinfo {pages} {1604310} (\bibinfo {year}
  {2017})}\BibitemShut {NoStop}%
\bibitem [{\citenamefont {Waser}\ \emph {et~al.}(2019)\citenamefont {Waser},
  \citenamefont {Dittmann}, \citenamefont {Menzel},\ and\ \citenamefont
  {Noll}}]{waser2019introduction}%
  \BibitemOpen
  \bibfield  {author} {\bibinfo {author} {\bibfnamefont {R.}~\bibnamefont
  {Waser}}, \bibinfo {author} {\bibfnamefont {R.}~\bibnamefont {Dittmann}},
  \bibinfo {author} {\bibfnamefont {S.}~\bibnamefont {Menzel}}, \ and\ \bibinfo
  {author} {\bibfnamefont {T.}~\bibnamefont {Noll}},\ }\href@noop {} {\bibfield
   {journal} {\bibinfo  {journal} {Faraday discussions}\ }\textbf {\bibinfo
  {volume} {213}},\ \bibinfo {pages} {11} (\bibinfo {year} {2019})}\BibitemShut
  {NoStop}%
\bibitem [{\citenamefont {Zafeiratos}\ \emph {et~al.}(2012)\citenamefont
  {Zafeiratos}, \citenamefont {Piccinin},\ and\ \citenamefont
  {Teschner}}]{zafeiratos2012alloys}%
  \BibitemOpen
  \bibfield  {author} {\bibinfo {author} {\bibfnamefont {S.}~\bibnamefont
  {Zafeiratos}}, \bibinfo {author} {\bibfnamefont {S.}~\bibnamefont
  {Piccinin}}, \ and\ \bibinfo {author} {\bibfnamefont {D.}~\bibnamefont
  {Teschner}},\ }\href@noop {} {\bibfield  {journal} {\bibinfo  {journal}
  {Catalysis Science \& Technology}\ }\textbf {\bibinfo {volume} {2}},\
  \bibinfo {pages} {1787} (\bibinfo {year} {2012})}\BibitemShut {NoStop}%
\bibitem [{\citenamefont {Hildebrand}\ \emph {et~al.}(1999)\citenamefont
  {Hildebrand}, \citenamefont {Kuperman}, \citenamefont {Wio}, \citenamefont
  {Mikhailov},\ and\ \citenamefont {Ertl}}]{hildebrand1999self}%
  \BibitemOpen
  \bibfield  {author} {\bibinfo {author} {\bibfnamefont {M.}~\bibnamefont
  {Hildebrand}}, \bibinfo {author} {\bibfnamefont {M.}~\bibnamefont
  {Kuperman}}, \bibinfo {author} {\bibfnamefont {H.}~\bibnamefont {Wio}},
  \bibinfo {author} {\bibfnamefont {A.}~\bibnamefont {Mikhailov}}, \ and\
  \bibinfo {author} {\bibfnamefont {G.}~\bibnamefont {Ertl}},\ }\href@noop {}
  {\bibfield  {journal} {\bibinfo  {journal} {Physical review letters}\
  }\textbf {\bibinfo {volume} {83}},\ \bibinfo {pages} {1475} (\bibinfo {year}
  {1999})}\BibitemShut {NoStop}%
\bibitem [{\citenamefont {Mikhailov}\ and\ \citenamefont
  {Ertl}(2009)}]{Mikhailov2009}%
  \BibitemOpen
  \bibfield  {author} {\bibinfo {author} {\bibfnamefont {A.~S.}\ \bibnamefont
  {Mikhailov}}\ and\ \bibinfo {author} {\bibfnamefont {G.}~\bibnamefont
  {Ertl}},\ }\href {\doibase 10.1002/cphc.200800277} {\bibfield  {journal}
  {\bibinfo  {journal} {ChemPhysChem}\ }\textbf {\bibinfo {volume} {10}},\
  \bibinfo {pages} {86} (\bibinfo {year} {2009})}\BibitemShut {NoStop}%
\bibitem [{\citenamefont {Nadkarni}\ \emph {et~al.}(2019)\citenamefont
  {Nadkarni}, \citenamefont {Zhou}, \citenamefont {Fraggedakis}, \citenamefont
  {Gao},\ and\ \citenamefont {Bazant}}]{nadkarni2019modeling}%
  \BibitemOpen
  \bibfield  {author} {\bibinfo {author} {\bibfnamefont {N.}~\bibnamefont
  {Nadkarni}}, \bibinfo {author} {\bibfnamefont {T.}~\bibnamefont {Zhou}},
  \bibinfo {author} {\bibfnamefont {D.}~\bibnamefont {Fraggedakis}}, \bibinfo
  {author} {\bibfnamefont {T.}~\bibnamefont {Gao}}, \ and\ \bibinfo {author}
  {\bibfnamefont {M.~Z.}\ \bibnamefont {Bazant}},\ }\href@noop {} {\bibfield
  {journal} {\bibinfo  {journal} {Advanced Functional Materials}\ }\textbf
  {\bibinfo {volume} {29}},\ \bibinfo {pages} {1902821} (\bibinfo {year}
  {2019})}\BibitemShut {NoStop}%
\bibitem [{\citenamefont {Gonzalez-Rosillo}\ \emph {et~al.}(2020)\citenamefont
  {Gonzalez-Rosillo}, \citenamefont {Balaish}, \citenamefont {Hood},
  \citenamefont {Nadkarni}, \citenamefont {Fraggedakis}, \citenamefont {Kim},
  \citenamefont {Mullin}, \citenamefont {Pfenninger}, \citenamefont {Bazant},\
  and\ \citenamefont {Rupp}}]{gonzalez2020lithium}%
  \BibitemOpen
  \bibfield  {author} {\bibinfo {author} {\bibfnamefont {J.~C.}\ \bibnamefont
  {Gonzalez-Rosillo}}, \bibinfo {author} {\bibfnamefont {M.}~\bibnamefont
  {Balaish}}, \bibinfo {author} {\bibfnamefont {Z.~D.}\ \bibnamefont {Hood}},
  \bibinfo {author} {\bibfnamefont {N.}~\bibnamefont {Nadkarni}}, \bibinfo
  {author} {\bibfnamefont {D.}~\bibnamefont {Fraggedakis}}, \bibinfo {author}
  {\bibfnamefont {K.~J.}\ \bibnamefont {Kim}}, \bibinfo {author} {\bibfnamefont
  {K.~M.}\ \bibnamefont {Mullin}}, \bibinfo {author} {\bibfnamefont
  {R.}~\bibnamefont {Pfenninger}}, \bibinfo {author} {\bibfnamefont {M.~Z.}\
  \bibnamefont {Bazant}}, \ and\ \bibinfo {author} {\bibfnamefont {J.~L.}\
  \bibnamefont {Rupp}},\ }\href@noop {} {\bibfield  {journal} {\bibinfo
  {journal} {Advanced Materials}\ }\textbf {\bibinfo {volume} {32}},\ \bibinfo
  {pages} {1907465} (\bibinfo {year} {2020})}\BibitemShut {NoStop}%
\bibitem [{\citenamefont {Lu}\ \emph {et~al.}(2020)\citenamefont {Lu},
  \citenamefont {Huberman}, \citenamefont {Zhang}, \citenamefont {Song},
  \citenamefont {Wang}, \citenamefont {Vardar}, \citenamefont {Hunt},
  \citenamefont {Waluyo}, \citenamefont {Chen},\ and\ \citenamefont
  {Yildiz}}]{lu2020bi}%
  \BibitemOpen
  \bibfield  {author} {\bibinfo {author} {\bibfnamefont {Q.}~\bibnamefont
  {Lu}}, \bibinfo {author} {\bibfnamefont {S.}~\bibnamefont {Huberman}},
  \bibinfo {author} {\bibfnamefont {H.}~\bibnamefont {Zhang}}, \bibinfo
  {author} {\bibfnamefont {Q.}~\bibnamefont {Song}}, \bibinfo {author}
  {\bibfnamefont {J.}~\bibnamefont {Wang}}, \bibinfo {author} {\bibfnamefont
  {G.}~\bibnamefont {Vardar}}, \bibinfo {author} {\bibfnamefont
  {A.}~\bibnamefont {Hunt}}, \bibinfo {author} {\bibfnamefont {I.}~\bibnamefont
  {Waluyo}}, \bibinfo {author} {\bibfnamefont {G.}~\bibnamefont {Chen}}, \ and\
  \bibinfo {author} {\bibfnamefont {B.}~\bibnamefont {Yildiz}},\ }\href@noop {}
  {\bibfield  {journal} {\bibinfo  {journal} {Nature Materials}\ ,\ \bibinfo
  {pages} {1}} (\bibinfo {year} {2020})}\BibitemShut {NoStop}%
\bibitem [{\citenamefont {T~O’Connor}\ \emph {et~al.}(2016)\citenamefont
  {T~O’Connor}, \citenamefont {Welland}, \citenamefont {Liu},\ and\
  \citenamefont {Voorhees}}]{t2016phase}%
  \BibitemOpen
  \bibfield  {author} {\bibinfo {author} {\bibfnamefont {D.}~\bibnamefont
  {T~O’Connor}}, \bibinfo {author} {\bibfnamefont {M.~J.}\ \bibnamefont
  {Welland}}, \bibinfo {author} {\bibfnamefont {W.~K.}\ \bibnamefont {Liu}}, \
  and\ \bibinfo {author} {\bibfnamefont {P.~W.}\ \bibnamefont {Voorhees}},\
  }\href@noop {} {\bibfield  {journal} {\bibinfo  {journal} {Modelling and
  Simulation in Materials Science and Engineering}\ }\textbf {\bibinfo {volume}
  {24}},\ \bibinfo {pages} {035020} (\bibinfo {year} {2016})}\BibitemShut
  {NoStop}%
\bibitem [{\citenamefont {Di~Leo}\ \emph {et~al.}(2015)\citenamefont {Di~Leo},
  \citenamefont {Rejovitzky},\ and\ \citenamefont {Anand}}]{di2015diffusion}%
  \BibitemOpen
  \bibfield  {author} {\bibinfo {author} {\bibfnamefont {C.~V.}\ \bibnamefont
  {Di~Leo}}, \bibinfo {author} {\bibfnamefont {E.}~\bibnamefont {Rejovitzky}},
  \ and\ \bibinfo {author} {\bibfnamefont {L.}~\bibnamefont {Anand}},\
  }\href@noop {} {\bibfield  {journal} {\bibinfo  {journal} {International
  Journal of Solids and Structures}\ }\textbf {\bibinfo {volume} {67}},\
  \bibinfo {pages} {283} (\bibinfo {year} {2015})}\BibitemShut {NoStop}%
\bibitem [{\citenamefont {Koerver}\ \emph {et~al.}(2018)\citenamefont
  {Koerver}, \citenamefont {Zhang}, \citenamefont {{De Biasi}}, \citenamefont
  {Schweidler}, \citenamefont {Kondrakov}, \citenamefont {Kolling},
  \citenamefont {Brezesinski}, \citenamefont {Hartmann}, \citenamefont
  {Zeier},\ and\ \citenamefont {Janek}}]{Koerver2018}%
  \BibitemOpen
  \bibfield  {author} {\bibinfo {author} {\bibfnamefont {R.}~\bibnamefont
  {Koerver}}, \bibinfo {author} {\bibfnamefont {W.}~\bibnamefont {Zhang}},
  \bibinfo {author} {\bibfnamefont {L.}~\bibnamefont {{De Biasi}}}, \bibinfo
  {author} {\bibfnamefont {S.}~\bibnamefont {Schweidler}}, \bibinfo {author}
  {\bibfnamefont {A.~O.}\ \bibnamefont {Kondrakov}}, \bibinfo {author}
  {\bibfnamefont {S.}~\bibnamefont {Kolling}}, \bibinfo {author} {\bibfnamefont
  {T.}~\bibnamefont {Brezesinski}}, \bibinfo {author} {\bibfnamefont
  {P.}~\bibnamefont {Hartmann}}, \bibinfo {author} {\bibfnamefont {W.~G.}\
  \bibnamefont {Zeier}}, \ and\ \bibinfo {author} {\bibfnamefont
  {J.}~\bibnamefont {Janek}},\ }\href {\doibase 10.1039/c8ee00907d} {\bibfield
  {journal} {\bibinfo  {journal} {Energy and Environmental Science}\ }\textbf
  {\bibinfo {volume} {11}},\ \bibinfo {pages} {2142} (\bibinfo {year}
  {2018})}\BibitemShut {NoStop}%
\bibitem [{\citenamefont {Kondepudi}\ and\ \citenamefont
  {Prigogine}(2014)}]{kondepudi2014modern}%
  \BibitemOpen
  \bibfield  {author} {\bibinfo {author} {\bibfnamefont {D.}~\bibnamefont
  {Kondepudi}}\ and\ \bibinfo {author} {\bibfnamefont {I.}~\bibnamefont
  {Prigogine}},\ }\href@noop {} {\emph {\bibinfo {title} {Modern
  thermodynamics: from heat engines to dissipative structures}}}\ (\bibinfo
  {publisher} {John Wiley \& Sons},\ \bibinfo {year} {2014})\BibitemShut
  {NoStop}%
\bibitem [{\citenamefont {Balluffi}\ \emph {et~al.}(2005)\citenamefont
  {Balluffi}, \citenamefont {Allen},\ and\ \citenamefont
  {Carter}}]{balluffi2005kinetics}%
  \BibitemOpen
  \bibfield  {author} {\bibinfo {author} {\bibfnamefont {R.~W.}\ \bibnamefont
  {Balluffi}}, \bibinfo {author} {\bibfnamefont {S.~M.}\ \bibnamefont {Allen}},
  \ and\ \bibinfo {author} {\bibfnamefont {W.~C.}\ \bibnamefont {Carter}},\
  }\href@noop {} {\emph {\bibinfo {title} {Kinetics of materials}}}\ (\bibinfo
  {publisher} {John Wiley \& Sons},\ \bibinfo {year} {2005})\BibitemShut
  {NoStop}%
\bibitem [{\citenamefont {Mirzadeh}\ and\ \citenamefont
  {Bazant}(2017)}]{mirzadeh2017electrokinetic}%
  \BibitemOpen
  \bibfield  {author} {\bibinfo {author} {\bibfnamefont {M.}~\bibnamefont
  {Mirzadeh}}\ and\ \bibinfo {author} {\bibfnamefont {M.~Z.}\ \bibnamefont
  {Bazant}},\ }\href@noop {} {\bibfield  {journal} {\bibinfo  {journal}
  {Physical review letters}\ }\textbf {\bibinfo {volume} {119}},\ \bibinfo
  {pages} {174501} (\bibinfo {year} {2017})}\BibitemShut {NoStop}%
\bibitem [{\citenamefont {Gao}\ \emph {et~al.}(2019)\citenamefont {Gao},
  \citenamefont {Mirzadeh}, \citenamefont {Bai}, \citenamefont {Conforti},\
  and\ \citenamefont {Bazant}}]{gao2019active}%
  \BibitemOpen
  \bibfield  {author} {\bibinfo {author} {\bibfnamefont {T.}~\bibnamefont
  {Gao}}, \bibinfo {author} {\bibfnamefont {M.}~\bibnamefont {Mirzadeh}},
  \bibinfo {author} {\bibfnamefont {P.}~\bibnamefont {Bai}}, \bibinfo {author}
  {\bibfnamefont {K.~M.}\ \bibnamefont {Conforti}}, \ and\ \bibinfo {author}
  {\bibfnamefont {M.~Z.}\ \bibnamefont {Bazant}},\ }\href@noop {} {\bibfield
  {journal} {\bibinfo  {journal} {Nature communications}\ }\textbf {\bibinfo
  {volume} {10}},\ \bibinfo {pages} {1} (\bibinfo {year} {2019})}\BibitemShut
  {NoStop}%
\bibitem [{\citenamefont {Bai}\ \emph {et~al.}(2011)\citenamefont {Bai},
  \citenamefont {Cogswell},\ and\ \citenamefont {Bazant}}]{bai2011suppression}%
  \BibitemOpen
  \bibfield  {author} {\bibinfo {author} {\bibfnamefont {P.}~\bibnamefont
  {Bai}}, \bibinfo {author} {\bibfnamefont {D.~A.}\ \bibnamefont {Cogswell}}, \
  and\ \bibinfo {author} {\bibfnamefont {M.~Z.}\ \bibnamefont {Bazant}},\
  }\href@noop {} {\bibfield  {journal} {\bibinfo  {journal} {Nano letters}\
  }\textbf {\bibinfo {volume} {11}},\ \bibinfo {pages} {4890} (\bibinfo {year}
  {2011})}\BibitemShut {NoStop}%
\bibitem [{\citenamefont {Bazant}(2017)}]{bazant2017thermodynamic}%
  \BibitemOpen
  \bibfield  {author} {\bibinfo {author} {\bibfnamefont {M.~Z.}\ \bibnamefont
  {Bazant}},\ }\href@noop {} {\bibfield  {journal} {\bibinfo  {journal}
  {Faraday discussions}\ }\textbf {\bibinfo {volume} {199}},\ \bibinfo {pages}
  {423} (\bibinfo {year} {2017})}\BibitemShut {NoStop}%
\bibitem [{\citenamefont {Fraggedakis}\ and\ \citenamefont
  {Bazant}(2020)}]{fraggedakis2020tuning}%
  \BibitemOpen
  \bibfield  {author} {\bibinfo {author} {\bibfnamefont {D.}~\bibnamefont
  {Fraggedakis}}\ and\ \bibinfo {author} {\bibfnamefont {M.~Z.}\ \bibnamefont
  {Bazant}},\ }\href@noop {} {\bibfield  {journal} {\bibinfo  {journal} {The
  Journal of Chemical Physics}\ }\textbf {\bibinfo {volume} {152}},\ \bibinfo
  {pages} {184703} (\bibinfo {year} {2020})}\BibitemShut {NoStop}%
\bibitem [{\citenamefont {Carmack}\ and\ \citenamefont
  {Millett}(2018)}]{carmack2018tuning}%
  \BibitemOpen
  \bibfield  {author} {\bibinfo {author} {\bibfnamefont {J.~M.}\ \bibnamefont
  {Carmack}}\ and\ \bibinfo {author} {\bibfnamefont {P.~C.}\ \bibnamefont
  {Millett}},\ }\href@noop {} {\bibfield  {journal} {\bibinfo  {journal} {Soft
  matter}\ }\textbf {\bibinfo {volume} {14}},\ \bibinfo {pages} {4344}
  (\bibinfo {year} {2018})}\BibitemShut {NoStop}%
\bibitem [{\citenamefont {Gu}\ \emph {et~al.}(2000)\citenamefont {Gu},
  \citenamefont {Liu},\ and\ \citenamefont {Liang}}]{gu2000effect}%
  \BibitemOpen
  \bibfield  {author} {\bibinfo {author} {\bibfnamefont {X.}~\bibnamefont
  {Gu}}, \bibinfo {author} {\bibfnamefont {W.}~\bibnamefont {Liu}}, \ and\
  \bibinfo {author} {\bibfnamefont {K.}~\bibnamefont {Liang}},\ }\href@noop {}
  {\bibfield  {journal} {\bibinfo  {journal} {Materials Science and
  Engineering: A}\ }\textbf {\bibinfo {volume} {278}},\ \bibinfo {pages} {22}
  (\bibinfo {year} {2000})}\BibitemShut {NoStop}%
\bibitem [{\citenamefont {Thornburg}\ and\ \citenamefont
  {White}(1972)}]{thornburg1972electric}%
  \BibitemOpen
  \bibfield  {author} {\bibinfo {author} {\bibfnamefont {D.~D.}\ \bibnamefont
  {Thornburg}}\ and\ \bibinfo {author} {\bibfnamefont {R.~M.}\ \bibnamefont
  {White}},\ }\href@noop {} {\bibfield  {journal} {\bibinfo  {journal} {Journal
  of Applied Physics}\ }\textbf {\bibinfo {volume} {43}},\ \bibinfo {pages}
  {4609} (\bibinfo {year} {1972})}\BibitemShut {NoStop}%
\bibitem [{\citenamefont {Hori}\ \emph {et~al.}(2007)\citenamefont {Hori},
  \citenamefont {Urakawa}, \citenamefont {Yano},\ and\ \citenamefont
  {Tran-Cong-Miyata}}]{hori2007phase}%
  \BibitemOpen
  \bibfield  {author} {\bibinfo {author} {\bibfnamefont {H.}~\bibnamefont
  {Hori}}, \bibinfo {author} {\bibfnamefont {O.}~\bibnamefont {Urakawa}},
  \bibinfo {author} {\bibfnamefont {O.}~\bibnamefont {Yano}}, \ and\ \bibinfo
  {author} {\bibfnamefont {Q.}~\bibnamefont {Tran-Cong-Miyata}},\ }\href@noop
  {} {\bibfield  {journal} {\bibinfo  {journal} {Macromolecules}\ }\textbf
  {\bibinfo {volume} {40}},\ \bibinfo {pages} {389} (\bibinfo {year}
  {2007})}\BibitemShut {NoStop}%
\bibitem [{\citenamefont {Aranson}\ \emph {et~al.}(2002)\citenamefont
  {Aranson}, \citenamefont {Meerson}, \citenamefont {Sasorov},\ and\
  \citenamefont {Vinokur}}]{aranson2002phase}%
  \BibitemOpen
  \bibfield  {author} {\bibinfo {author} {\bibfnamefont {I.}~\bibnamefont
  {Aranson}}, \bibinfo {author} {\bibfnamefont {B.}~\bibnamefont {Meerson}},
  \bibinfo {author} {\bibfnamefont {P.}~\bibnamefont {Sasorov}}, \ and\
  \bibinfo {author} {\bibfnamefont {V.}~\bibnamefont {Vinokur}},\ }\href@noop
  {} {\bibfield  {journal} {\bibinfo  {journal} {Physical review letters}\
  }\textbf {\bibinfo {volume} {88}},\ \bibinfo {pages} {204301} (\bibinfo
  {year} {2002})}\BibitemShut {NoStop}%
\bibitem [{\citenamefont {Tsori}\ and\ \citenamefont
  {Leibler}(2007)}]{Tsori2007}%
  \BibitemOpen
  \bibfield  {author} {\bibinfo {author} {\bibfnamefont {Y.}~\bibnamefont
  {Tsori}}\ and\ \bibinfo {author} {\bibfnamefont {L.}~\bibnamefont
  {Leibler}},\ }\href {\doibase 10.1073/pnas.0607746104} {\bibfield  {journal}
  {\bibinfo  {journal} {Proceedings of the National Academy of Sciences of the
  United States of America}\ }\textbf {\bibinfo {volume} {104}},\ \bibinfo
  {pages} {7348} (\bibinfo {year} {2007})}\BibitemShut {NoStop}%
\bibitem [{\citenamefont {Tsori}\ \emph {et~al.}(2004)\citenamefont {Tsori},
  \citenamefont {Tournilhac},\ and\ \citenamefont {Leibler}}]{Tsori2004}%
  \BibitemOpen
  \bibfield  {author} {\bibinfo {author} {\bibfnamefont {Y.}~\bibnamefont
  {Tsori}}, \bibinfo {author} {\bibfnamefont {F.}~\bibnamefont {Tournilhac}}, \
  and\ \bibinfo {author} {\bibfnamefont {L.}~\bibnamefont {Leibler}},\ }\href
  {\doibase 10.1038/nature02758} {\bibfield  {journal} {\bibinfo  {journal}
  {Nature}\ }\textbf {\bibinfo {volume} {430}},\ \bibinfo {pages} {544}
  (\bibinfo {year} {2004})}\BibitemShut {NoStop}%
\bibitem [{\citenamefont {Wang}(2008)}]{wang2008effects}%
  \BibitemOpen
  \bibfield  {author} {\bibinfo {author} {\bibfnamefont {Z.-G.}\ \bibnamefont
  {Wang}},\ }\href@noop {} {\bibfield  {journal} {\bibinfo  {journal} {The
  Journal of Physical Chemistry B}\ }\textbf {\bibinfo {volume} {112}},\
  \bibinfo {pages} {16205} (\bibinfo {year} {2008})}\BibitemShut {NoStop}%
\bibitem [{\citenamefont {Khusid}\ and\ \citenamefont
  {Acrivos}(1996)}]{khusid1996effects}%
  \BibitemOpen
  \bibfield  {author} {\bibinfo {author} {\bibfnamefont {B.}~\bibnamefont
  {Khusid}}\ and\ \bibinfo {author} {\bibfnamefont {A.}~\bibnamefont
  {Acrivos}},\ }\href@noop {} {\bibfield  {journal} {\bibinfo  {journal}
  {Physical Review E}\ }\textbf {\bibinfo {volume} {54}},\ \bibinfo {pages}
  {5428} (\bibinfo {year} {1996})}\BibitemShut {NoStop}%
\bibitem [{\citenamefont {Khusid}\ and\ \citenamefont
  {Acrivos}(1999)}]{khusid1999phase}%
  \BibitemOpen
  \bibfield  {author} {\bibinfo {author} {\bibfnamefont {B.}~\bibnamefont
  {Khusid}}\ and\ \bibinfo {author} {\bibfnamefont {A.}~\bibnamefont
  {Acrivos}},\ }\href@noop {} {\bibfield  {journal} {\bibinfo  {journal}
  {Physical Review E}\ }\textbf {\bibinfo {volume} {60}},\ \bibinfo {pages}
  {3015} (\bibinfo {year} {1999})}\BibitemShut {NoStop}%
\bibitem [{\citenamefont {Kumar}\ \emph {et~al.}(2004)\citenamefont {Kumar},
  \citenamefont {Qiu}, \citenamefont {Acrivos}, \citenamefont {Khusid},\ and\
  \citenamefont {Jacqmin}}]{kumar2004combined}%
  \BibitemOpen
  \bibfield  {author} {\bibinfo {author} {\bibfnamefont {A.}~\bibnamefont
  {Kumar}}, \bibinfo {author} {\bibfnamefont {Z.}~\bibnamefont {Qiu}}, \bibinfo
  {author} {\bibfnamefont {A.}~\bibnamefont {Acrivos}}, \bibinfo {author}
  {\bibfnamefont {B.}~\bibnamefont {Khusid}}, \ and\ \bibinfo {author}
  {\bibfnamefont {D.}~\bibnamefont {Jacqmin}},\ }\href@noop {} {\bibfield
  {journal} {\bibinfo  {journal} {Physical Review E}\ }\textbf {\bibinfo
  {volume} {69}},\ \bibinfo {pages} {021402} (\bibinfo {year}
  {2004})}\BibitemShut {NoStop}%
\bibitem [{\citenamefont {Johnson}\ \emph {et~al.}(2004)\citenamefont
  {Johnson}, \citenamefont {Duan}, \citenamefont {Riley}, \citenamefont
  {Bhattacharya},\ and\ \citenamefont {Luo}}]{Johnson2004}%
  \BibitemOpen
  \bibfield  {author} {\bibinfo {author} {\bibfnamefont {M.~D.}\ \bibnamefont
  {Johnson}}, \bibinfo {author} {\bibfnamefont {X.}~\bibnamefont {Duan}},
  \bibinfo {author} {\bibfnamefont {B.}~\bibnamefont {Riley}}, \bibinfo
  {author} {\bibfnamefont {A.}~\bibnamefont {Bhattacharya}}, \ and\ \bibinfo
  {author} {\bibfnamefont {W.}~\bibnamefont {Luo}},\ }\href {\doibase
  10.1103/PhysRevE.69.041501} {\bibfield  {journal} {\bibinfo  {journal}
  {Physical Review E - Statistical Physics, Plasmas, Fluids, and Related
  Interdisciplinary Topics}\ }\textbf {\bibinfo {volume} {69}},\ \bibinfo
  {pages} {7} (\bibinfo {year} {2004})}\BibitemShut {NoStop}%
\bibitem [{\citenamefont {Liedel}\ \emph
  {et~al.}(2012{\natexlab{a}})\citenamefont {Liedel}, \citenamefont {Pester},
  \citenamefont {Ruppel}, \citenamefont {Urban},\ and\ \citenamefont
  {B{\"{o}}ker}}]{Liedel2012}%
  \BibitemOpen
  \bibfield  {author} {\bibinfo {author} {\bibfnamefont {C.}~\bibnamefont
  {Liedel}}, \bibinfo {author} {\bibfnamefont {C.~W.}\ \bibnamefont {Pester}},
  \bibinfo {author} {\bibfnamefont {M.}~\bibnamefont {Ruppel}}, \bibinfo
  {author} {\bibfnamefont {V.~S.}\ \bibnamefont {Urban}}, \ and\ \bibinfo
  {author} {\bibfnamefont {A.}~\bibnamefont {B{\"{o}}ker}},\ }\href {\doibase
  10.1002/macp.201100590} {\bibfield  {journal} {\bibinfo  {journal}
  {Macromolecular Chemistry and Physics}\ }\textbf {\bibinfo {volume} {213}},\
  \bibinfo {pages} {259} (\bibinfo {year} {2012}{\natexlab{a}})}\BibitemShut
  {NoStop}%
\bibitem [{\citenamefont {Tomlinson}(1972)}]{tomlinson1972spinodal}%
  \BibitemOpen
  \bibfield  {author} {\bibinfo {author} {\bibfnamefont {J.}~\bibnamefont
  {Tomlinson}},\ }\href@noop {} {\bibfield  {journal} {\bibinfo  {journal}
  {Journal of Electronic Materials}\ }\textbf {\bibinfo {volume} {1}},\
  \bibinfo {pages} {357} (\bibinfo {year} {1972})}\BibitemShut {NoStop}%
\bibitem [{\citenamefont {Bazant}(2013)}]{bazant2013theory}%
  \BibitemOpen
  \bibfield  {author} {\bibinfo {author} {\bibfnamefont {M.~Z.}\ \bibnamefont
  {Bazant}},\ }\href@noop {} {\bibfield  {journal} {\bibinfo  {journal}
  {Accounts of chemical research}\ }\textbf {\bibinfo {volume} {46}},\ \bibinfo
  {pages} {1144} (\bibinfo {year} {2013})}\BibitemShut {NoStop}%
\bibitem [{\citenamefont {Kupershtokh}\ and\ \citenamefont
  {Medvedev}(2006)}]{kupershtokh2006anisotropic}%
  \BibitemOpen
  \bibfield  {author} {\bibinfo {author} {\bibfnamefont {A.~L.}\ \bibnamefont
  {Kupershtokh}}\ and\ \bibinfo {author} {\bibfnamefont {D.~A.}\ \bibnamefont
  {Medvedev}},\ }\href@noop {} {\bibfield  {journal} {\bibinfo  {journal}
  {Physical Review E}\ }\textbf {\bibinfo {volume} {74}},\ \bibinfo {pages}
  {021505} (\bibinfo {year} {2006})}\BibitemShut {NoStop}%
\bibitem [{\citenamefont {Liedel}\ \emph
  {et~al.}(2012{\natexlab{b}})\citenamefont {Liedel}, \citenamefont {Pester},
  \citenamefont {Ruppel}, \citenamefont {Urban},\ and\ \citenamefont
  {B{\"o}ker}}]{liedel2012beyond}%
  \BibitemOpen
  \bibfield  {author} {\bibinfo {author} {\bibfnamefont {C.}~\bibnamefont
  {Liedel}}, \bibinfo {author} {\bibfnamefont {C.~W.}\ \bibnamefont {Pester}},
  \bibinfo {author} {\bibfnamefont {M.}~\bibnamefont {Ruppel}}, \bibinfo
  {author} {\bibfnamefont {V.~S.}\ \bibnamefont {Urban}}, \ and\ \bibinfo
  {author} {\bibfnamefont {A.}~\bibnamefont {B{\"o}ker}},\ }\href@noop {}
  {\bibfield  {journal} {\bibinfo  {journal} {Macromolecular Chemistry and
  Physics}\ }\textbf {\bibinfo {volume} {213}},\ \bibinfo {pages} {259}
  (\bibinfo {year} {2012}{\natexlab{b}})}\BibitemShut {NoStop}%
\bibitem [{\citenamefont {Sze}\ and\ \citenamefont
  {Ng}(2006)}]{sze2006physics}%
  \BibitemOpen
  \bibfield  {author} {\bibinfo {author} {\bibfnamefont {S.~M.}\ \bibnamefont
  {Sze}}\ and\ \bibinfo {author} {\bibfnamefont {K.~K.}\ \bibnamefont {Ng}},\
  }\href@noop {} {\emph {\bibinfo {title} {Physics of semiconductor devices}}}\
  (\bibinfo  {publisher} {John wiley \& sons},\ \bibinfo {year}
  {2006})\BibitemShut {NoStop}%
\bibitem [{\citenamefont {O'Dwyer}(1973)}]{o1973theory}%
  \BibitemOpen
  \bibfield  {author} {\bibinfo {author} {\bibfnamefont {J.~J.}\ \bibnamefont
  {O'Dwyer}},\ }\href@noop {} {\emph {\bibinfo {title} {The theory of
  electrical conduction and breakdown in solid dielectrics}}}\ (\bibinfo
  {publisher} {Clarendon Press},\ \bibinfo {year} {1973})\BibitemShut {NoStop}%
\bibitem [{\citenamefont {Milewska}\ \emph {et~al.}(2014)\citenamefont
  {Milewska}, \citenamefont {{\'S}wierczek}, \citenamefont {Tobola},
  \citenamefont {Boudoire}, \citenamefont {Hu}, \citenamefont {Bora},
  \citenamefont {Mun}, \citenamefont {Braun},\ and\ \citenamefont
  {Molenda}}]{milewska2014nature}%
  \BibitemOpen
  \bibfield  {author} {\bibinfo {author} {\bibfnamefont {A.}~\bibnamefont
  {Milewska}}, \bibinfo {author} {\bibfnamefont {K.}~\bibnamefont
  {{\'S}wierczek}}, \bibinfo {author} {\bibfnamefont {J.}~\bibnamefont
  {Tobola}}, \bibinfo {author} {\bibfnamefont {F.}~\bibnamefont {Boudoire}},
  \bibinfo {author} {\bibfnamefont {Y.}~\bibnamefont {Hu}}, \bibinfo {author}
  {\bibfnamefont {D.}~\bibnamefont {Bora}}, \bibinfo {author} {\bibfnamefont
  {B.}~\bibnamefont {Mun}}, \bibinfo {author} {\bibfnamefont {A.}~\bibnamefont
  {Braun}}, \ and\ \bibinfo {author} {\bibfnamefont {J.}~\bibnamefont
  {Molenda}},\ }\href@noop {} {\bibfield  {journal} {\bibinfo  {journal} {Solid
  State Ionics}\ }\textbf {\bibinfo {volume} {263}},\ \bibinfo {pages} {110}
  (\bibinfo {year} {2014})}\BibitemShut {NoStop}%
\bibitem [{\citenamefont {Amin}\ and\ \citenamefont
  {Chiang}(2016)}]{amin2016characterization}%
  \BibitemOpen
  \bibfield  {author} {\bibinfo {author} {\bibfnamefont {R.}~\bibnamefont
  {Amin}}\ and\ \bibinfo {author} {\bibfnamefont {Y.-M.}\ \bibnamefont
  {Chiang}},\ }\href@noop {} {\bibfield  {journal} {\bibinfo  {journal}
  {Journal of The Electrochemical Society}\ }\textbf {\bibinfo {volume}
  {163}},\ \bibinfo {pages} {A1512} (\bibinfo {year} {2016})}\BibitemShut
  {NoStop}%
\bibitem [{\citenamefont {Verde}\ \emph {et~al.}(2016)\citenamefont {Verde},
  \citenamefont {Baggetto}, \citenamefont {Balke}, \citenamefont {Veith},
  \citenamefont {Seo}, \citenamefont {Wang},\ and\ \citenamefont
  {Meng}}]{verde2016elucidating}%
  \BibitemOpen
  \bibfield  {author} {\bibinfo {author} {\bibfnamefont {M.~G.}\ \bibnamefont
  {Verde}}, \bibinfo {author} {\bibfnamefont {L.}~\bibnamefont {Baggetto}},
  \bibinfo {author} {\bibfnamefont {N.}~\bibnamefont {Balke}}, \bibinfo
  {author} {\bibfnamefont {G.~M.}\ \bibnamefont {Veith}}, \bibinfo {author}
  {\bibfnamefont {J.~K.}\ \bibnamefont {Seo}}, \bibinfo {author} {\bibfnamefont
  {Z.}~\bibnamefont {Wang}}, \ and\ \bibinfo {author} {\bibfnamefont {Y.~S.}\
  \bibnamefont {Meng}},\ }\href@noop {} {\bibfield  {journal} {\bibinfo
  {journal} {ACS nano}\ }\textbf {\bibinfo {volume} {10}},\ \bibinfo {pages}
  {4312} (\bibinfo {year} {2016})}\BibitemShut {NoStop}%
\bibitem [{\citenamefont {Young}\ \emph {et~al.}(2013)\citenamefont {Young},
  \citenamefont {Ransil}, \citenamefont {Amin}, \citenamefont {Li},\ and\
  \citenamefont {Chiang}}]{young2013electronic}%
  \BibitemOpen
  \bibfield  {author} {\bibinfo {author} {\bibfnamefont {D.}~\bibnamefont
  {Young}}, \bibinfo {author} {\bibfnamefont {A.}~\bibnamefont {Ransil}},
  \bibinfo {author} {\bibfnamefont {R.}~\bibnamefont {Amin}}, \bibinfo {author}
  {\bibfnamefont {Z.}~\bibnamefont {Li}}, \ and\ \bibinfo {author}
  {\bibfnamefont {Y.-M.}\ \bibnamefont {Chiang}},\ }\href@noop {} {\bibfield
  {journal} {\bibinfo  {journal} {Advanced energy materials}\ }\textbf
  {\bibinfo {volume} {3}},\ \bibinfo {pages} {1125} (\bibinfo {year}
  {2013})}\BibitemShut {NoStop}%
\bibitem [{\citenamefont {Goldhammer}(1913)}]{goldhammer1913dispersion}%
  \BibitemOpen
  \bibfield  {author} {\bibinfo {author} {\bibfnamefont {D.~A.}\ \bibnamefont
  {Goldhammer}},\ }\href@noop {} {\emph {\bibinfo {title} {Dispersion und
  Absorption des Lichtes in ruhenden isotropen K{\"o}rpern: Theorie und ihre
  Folgerungen}}},\ Vol.~\bibinfo {volume} {16}\ (\bibinfo  {publisher}
  {Teubner},\ \bibinfo {year} {1913})\BibitemShut {NoStop}%
\bibitem [{\citenamefont {Herzfeld}(1927)}]{herzfeld1927atomic}%
  \BibitemOpen
  \bibfield  {author} {\bibinfo {author} {\bibfnamefont {K.}~\bibnamefont
  {Herzfeld}},\ }\href@noop {} {\bibfield  {journal} {\bibinfo  {journal}
  {Physical Review}\ }\textbf {\bibinfo {volume} {29}},\ \bibinfo {pages} {701}
  (\bibinfo {year} {1927})}\BibitemShut {NoStop}%
\bibitem [{\citenamefont {Manthiram}(2020)}]{manthiram2020reflection}%
  \BibitemOpen
  \bibfield  {author} {\bibinfo {author} {\bibfnamefont {A.}~\bibnamefont
  {Manthiram}},\ }\href@noop {} {\bibfield  {journal} {\bibinfo  {journal}
  {Nature Communications}\ }\textbf {\bibinfo {volume} {11}},\ \bibinfo {pages}
  {1} (\bibinfo {year} {2020})}\BibitemShut {NoStop}%
\bibitem [{\citenamefont {Nitta}\ \emph {et~al.}(2015)\citenamefont {Nitta},
  \citenamefont {Wu}, \citenamefont {Lee},\ and\ \citenamefont
  {Yushin}}]{Nitta2015}%
  \BibitemOpen
  \bibfield  {author} {\bibinfo {author} {\bibfnamefont {N.}~\bibnamefont
  {Nitta}}, \bibinfo {author} {\bibfnamefont {F.}~\bibnamefont {Wu}}, \bibinfo
  {author} {\bibfnamefont {J.~T.}\ \bibnamefont {Lee}}, \ and\ \bibinfo
  {author} {\bibfnamefont {G.}~\bibnamefont {Yushin}},\ }\href {\doibase
  10.1016/j.mattod.2014.10.040} {\bibfield  {journal} {\bibinfo  {journal}
  {Materials Today}\ }\textbf {\bibinfo {volume} {18}},\ \bibinfo {pages} {252}
  (\bibinfo {year} {2015})},\ \Eprint {http://arxiv.org/abs/arXiv:1011.1669v3}
  {arXiv:arXiv:1011.1669v3} \BibitemShut {NoStop}%
\bibitem [{\citenamefont {Li}\ \emph {et~al.}(2018)\citenamefont {Li},
  \citenamefont {Chen}, \citenamefont {Lim}, \citenamefont {Deng},
  \citenamefont {Lim}, \citenamefont {Fraggedakis}, \citenamefont {Attia},
  \citenamefont {Lee}, \citenamefont {Jin}, \citenamefont {Mo{\v{s}}kon},
  \citenamefont {Guan}, \citenamefont {Gent}, \citenamefont {Hong},
  \citenamefont {Yu}, \citenamefont {Gaber{\v{s}}{\v{c}}ek}, \citenamefont
  {Islam}, \citenamefont {Bazant},\ and\ \citenamefont {Chueh}}]{li2018fluid}%
  \BibitemOpen
  \bibfield  {author} {\bibinfo {author} {\bibfnamefont {Y.}~\bibnamefont
  {Li}}, \bibinfo {author} {\bibfnamefont {H.}~\bibnamefont {Chen}}, \bibinfo
  {author} {\bibfnamefont {K.}~\bibnamefont {Lim}}, \bibinfo {author}
  {\bibfnamefont {H.~D.}\ \bibnamefont {Deng}}, \bibinfo {author}
  {\bibfnamefont {J.}~\bibnamefont {Lim}}, \bibinfo {author} {\bibfnamefont
  {D.}~\bibnamefont {Fraggedakis}}, \bibinfo {author} {\bibfnamefont {P.~M.}\
  \bibnamefont {Attia}}, \bibinfo {author} {\bibfnamefont {S.~C.}\ \bibnamefont
  {Lee}}, \bibinfo {author} {\bibfnamefont {N.}~\bibnamefont {Jin}}, \bibinfo
  {author} {\bibfnamefont {J.}~\bibnamefont {Mo{\v{s}}kon}}, \bibinfo {author}
  {\bibfnamefont {Z.}~\bibnamefont {Guan}}, \bibinfo {author} {\bibfnamefont
  {W.~E.}\ \bibnamefont {Gent}}, \bibinfo {author} {\bibfnamefont
  {J.}~\bibnamefont {Hong}}, \bibinfo {author} {\bibfnamefont {Y.~S.}\
  \bibnamefont {Yu}}, \bibinfo {author} {\bibfnamefont {M.}~\bibnamefont
  {Gaber{\v{s}}{\v{c}}ek}}, \bibinfo {author} {\bibfnamefont {M.~S.}\
  \bibnamefont {Islam}}, \bibinfo {author} {\bibfnamefont {M.~Z.}\ \bibnamefont
  {Bazant}}, \ and\ \bibinfo {author} {\bibfnamefont {W.~C.}\ \bibnamefont
  {Chueh}},\ }\href@noop {} {\bibfield  {journal} {\bibinfo  {journal} {Nature
  materials}\ }\textbf {\bibinfo {volume} {17}},\ \bibinfo {pages} {915}
  (\bibinfo {year} {2018})}\BibitemShut {NoStop}%
\bibitem [{\citenamefont {Nadkarni}\ \emph {et~al.}(2018)\citenamefont
  {Nadkarni}, \citenamefont {Rejovitsky}, \citenamefont {Fraggedakis},
  \citenamefont {Di~Leo}, \citenamefont {Smith}, \citenamefont {Bai},\ and\
  \citenamefont {Bazant}}]{nadkarni2018interplay}%
  \BibitemOpen
  \bibfield  {author} {\bibinfo {author} {\bibfnamefont {N.}~\bibnamefont
  {Nadkarni}}, \bibinfo {author} {\bibfnamefont {E.}~\bibnamefont
  {Rejovitsky}}, \bibinfo {author} {\bibfnamefont {D.}~\bibnamefont
  {Fraggedakis}}, \bibinfo {author} {\bibfnamefont {C.~V.}\ \bibnamefont
  {Di~Leo}}, \bibinfo {author} {\bibfnamefont {R.~B.}\ \bibnamefont {Smith}},
  \bibinfo {author} {\bibfnamefont {P.}~\bibnamefont {Bai}}, \ and\ \bibinfo
  {author} {\bibfnamefont {M.~Z.}\ \bibnamefont {Bazant}},\ }\href@noop {}
  {\bibfield  {journal} {\bibinfo  {journal} {Physical Review Materials}\
  }\textbf {\bibinfo {volume} {2}},\ \bibinfo {pages} {085406} (\bibinfo {year}
  {2018})}\BibitemShut {NoStop}%
\bibitem [{\citenamefont {de~Klerk}\ \emph {et~al.}(2017)\citenamefont
  {de~Klerk}, \citenamefont {Vasileiadis}, \citenamefont {Smith}, \citenamefont
  {Bazant},\ and\ \citenamefont {Wagemaker}}]{de2017explaining}%
  \BibitemOpen
  \bibfield  {author} {\bibinfo {author} {\bibfnamefont {N.~J.}\ \bibnamefont
  {de~Klerk}}, \bibinfo {author} {\bibfnamefont {A.}~\bibnamefont
  {Vasileiadis}}, \bibinfo {author} {\bibfnamefont {R.~B.}\ \bibnamefont
  {Smith}}, \bibinfo {author} {\bibfnamefont {M.~Z.}\ \bibnamefont {Bazant}}, \
  and\ \bibinfo {author} {\bibfnamefont {M.}~\bibnamefont {Wagemaker}},\
  }\href@noop {} {\bibfield  {journal} {\bibinfo  {journal} {Physical Review
  Materials}\ }\textbf {\bibinfo {volume} {1}},\ \bibinfo {pages} {025404}
  (\bibinfo {year} {2017})}\BibitemShut {NoStop}%
\bibitem [{\citenamefont {Li}\ \emph {et~al.}(2019)\citenamefont {Li},
  \citenamefont {Fuller}, \citenamefont {Asapu}, \citenamefont {Agarwal},
  \citenamefont {Kurita}, \citenamefont {Yang},\ and\ \citenamefont
  {Talin}}]{li2019low}%
  \BibitemOpen
  \bibfield  {author} {\bibinfo {author} {\bibfnamefont {Y.}~\bibnamefont
  {Li}}, \bibinfo {author} {\bibfnamefont {E.~J.}\ \bibnamefont {Fuller}},
  \bibinfo {author} {\bibfnamefont {S.}~\bibnamefont {Asapu}}, \bibinfo
  {author} {\bibfnamefont {S.}~\bibnamefont {Agarwal}}, \bibinfo {author}
  {\bibfnamefont {T.}~\bibnamefont {Kurita}}, \bibinfo {author} {\bibfnamefont
  {J.~J.}\ \bibnamefont {Yang}}, \ and\ \bibinfo {author} {\bibfnamefont
  {A.~A.}\ \bibnamefont {Talin}},\ }\href@noop {} {\bibfield  {journal}
  {\bibinfo  {journal} {ACS Applied Materials \& Interfaces}\ }\textbf
  {\bibinfo {volume} {11}},\ \bibinfo {pages} {38982} (\bibinfo {year}
  {2019})}\BibitemShut {NoStop}%
\bibitem [{\citenamefont {Chaikin}\ \emph {et~al.}(1995)\citenamefont
  {Chaikin}, \citenamefont {Lubensky},\ and\ \citenamefont
  {Witten}}]{chaikin1995principles}%
  \BibitemOpen
  \bibfield  {author} {\bibinfo {author} {\bibfnamefont {P.~M.}\ \bibnamefont
  {Chaikin}}, \bibinfo {author} {\bibfnamefont {T.~C.}\ \bibnamefont
  {Lubensky}}, \ and\ \bibinfo {author} {\bibfnamefont {T.~A.}\ \bibnamefont
  {Witten}},\ }\href@noop {} {\emph {\bibinfo {title} {Principles of condensed
  matter physics}}},\ Vol.~\bibinfo {volume} {10}\ (\bibinfo  {publisher}
  {Cambridge university press Cambridge},\ \bibinfo {year} {1995})\BibitemShut
  {NoStop}%
\bibitem [{\citenamefont {van~der Waals}(1979)}]{van1979thermodynamic}%
  \BibitemOpen
  \bibfield  {author} {\bibinfo {author} {\bibfnamefont {J.~D.}\ \bibnamefont
  {van~der Waals}},\ }\href@noop {} {\bibfield  {journal} {\bibinfo  {journal}
  {Journal of Statistical Physics}\ }\textbf {\bibinfo {volume} {20}},\
  \bibinfo {pages} {200} (\bibinfo {year} {1979})}\BibitemShut {NoStop}%
\bibitem [{\citenamefont {Cahn}\ and\ \citenamefont
  {Hilliard}(1959)}]{cahn1959free}%
  \BibitemOpen
  \bibfield  {author} {\bibinfo {author} {\bibfnamefont {J.~W.}\ \bibnamefont
  {Cahn}}\ and\ \bibinfo {author} {\bibfnamefont {J.~E.}\ \bibnamefont
  {Hilliard}},\ }\href@noop {} {\bibfield  {journal} {\bibinfo  {journal} {The
  Journal of chemical physics}\ }\textbf {\bibinfo {volume} {31}},\ \bibinfo
  {pages} {688} (\bibinfo {year} {1959})}\BibitemShut {NoStop}%
\bibitem [{\citenamefont {Cahn}\ and\ \citenamefont
  {Hilliard}(1958)}]{cahn1958free}%
  \BibitemOpen
  \bibfield  {author} {\bibinfo {author} {\bibfnamefont {J.~W.}\ \bibnamefont
  {Cahn}}\ and\ \bibinfo {author} {\bibfnamefont {J.~E.}\ \bibnamefont
  {Hilliard}},\ }\href@noop {} {\bibfield  {journal} {\bibinfo  {journal} {The
  Journal of chemical physics}\ }\textbf {\bibinfo {volume} {28}},\ \bibinfo
  {pages} {258} (\bibinfo {year} {1958})}\BibitemShut {NoStop}%
\bibitem [{\citenamefont {Hildebrand}\ \emph {et~al.}(1936)\citenamefont
  {Hildebrand}, ,\ and\ \citenamefont {Scott}}]{hildebrand1936solubility}%
  \BibitemOpen
  \bibfield  {author} {\bibinfo {author} {\bibfnamefont {J.~H.}\ \bibnamefont
  {Hildebrand}}, , \ and\ \bibinfo {author} {\bibfnamefont {R.}~\bibnamefont
  {Scott}},\ }\href@noop {} {\emph {\bibinfo {title} {Solubility of
  Non-electrolytes}}}\ (\bibinfo  {publisher} {Rheinhold Publishing Co., New
  York},\ \bibinfo {year} {1936})\BibitemShut {NoStop}%
\bibitem [{\citenamefont {Zhou}\ \emph {et~al.}(2019)\citenamefont {Zhou},
  \citenamefont {Mirzadeh}, \citenamefont {Fraggedakis}, \citenamefont
  {Pellenq},\ and\ \citenamefont {Bazant}}]{zhou2019theory}%
  \BibitemOpen
  \bibfield  {author} {\bibinfo {author} {\bibfnamefont {T.}~\bibnamefont
  {Zhou}}, \bibinfo {author} {\bibfnamefont {M.}~\bibnamefont {Mirzadeh}},
  \bibinfo {author} {\bibfnamefont {D.}~\bibnamefont {Fraggedakis}}, \bibinfo
  {author} {\bibfnamefont {R.~J.-M.}\ \bibnamefont {Pellenq}}, \ and\ \bibinfo
  {author} {\bibfnamefont {M.~Z.}\ \bibnamefont {Bazant}},\ }\href@noop {}
  {\bibfield  {journal} {\bibinfo  {journal} {arXiv preprint arXiv:1909.09332}\
  } (\bibinfo {year} {2019})}\BibitemShut {NoStop}%
\bibitem [{\citenamefont {Gelfand}\ \emph {et~al.}(2000)\citenamefont
  {Gelfand}, \citenamefont {Silverman},\ and\ \citenamefont
  {Fomin}}]{gelfand2000calculus}%
  \BibitemOpen
  \bibfield  {author} {\bibinfo {author} {\bibfnamefont {I.~M.}\ \bibnamefont
  {Gelfand}}, \bibinfo {author} {\bibfnamefont {R.~A.}\ \bibnamefont
  {Silverman}}, \ and\ \bibinfo {author} {\bibfnamefont {S.}~\bibnamefont
  {Fomin}},\ }\href@noop {} {\emph {\bibinfo {title} {Calculus of
  variations}}}\ (\bibinfo  {publisher} {Courier Corporation},\ \bibinfo {year}
  {2000})\BibitemShut {NoStop}%
\bibitem [{\citenamefont {Brosseau}(1994)}]{brosseau1994dielectric}%
  \BibitemOpen
  \bibfield  {author} {\bibinfo {author} {\bibfnamefont {C.}~\bibnamefont
  {Brosseau}},\ }\href@noop {} {\bibfield  {journal} {\bibinfo  {journal}
  {Journal of applied physics}\ }\textbf {\bibinfo {volume} {75}},\ \bibinfo
  {pages} {672} (\bibinfo {year} {1994})}\BibitemShut {NoStop}%
\bibitem [{\citenamefont {Deen}(1998)}]{deen1998analysis}%
  \BibitemOpen
  \bibfield  {author} {\bibinfo {author} {\bibfnamefont {W.~M.}\ \bibnamefont
  {Deen}},\ }\href@noop {} {\emph {\bibinfo {title} {Analysis of transport
  phenomena}}}\ (\bibinfo  {publisher} {Oxford University Press New York},\
  \bibinfo {year} {1998})\BibitemShut {NoStop}%
\bibitem [{\citenamefont {Jackson}(2007)}]{jackson2007classical}%
  \BibitemOpen
  \bibfield  {author} {\bibinfo {author} {\bibfnamefont {J.~D.}\ \bibnamefont
  {Jackson}},\ }\href@noop {} {\emph {\bibinfo {title} {Classical
  electrodynamics}}}\ (\bibinfo  {publisher} {John Wiley \& Sons},\ \bibinfo
  {year} {2007})\BibitemShut {NoStop}%
\bibitem [{\citenamefont {De~Groot}\ and\ \citenamefont
  {Mazur}(2013)}]{de2013non}%
  \BibitemOpen
  \bibfield  {author} {\bibinfo {author} {\bibfnamefont {S.~R.}\ \bibnamefont
  {De~Groot}}\ and\ \bibinfo {author} {\bibfnamefont {P.}~\bibnamefont
  {Mazur}},\ }\href@noop {} {\emph {\bibinfo {title} {Non-equilibrium
  thermodynamics}}}\ (\bibinfo  {publisher} {Courier Corporation},\ \bibinfo
  {year} {2013})\BibitemShut {NoStop}%
\bibitem [{\citenamefont
  {Onsager}(1931{\natexlab{a}})}]{onsager1931reciprocal_1}%
  \BibitemOpen
  \bibfield  {author} {\bibinfo {author} {\bibfnamefont {L.}~\bibnamefont
  {Onsager}},\ }\href@noop {} {\bibfield  {journal} {\bibinfo  {journal}
  {Physical review}\ }\textbf {\bibinfo {volume} {37}},\ \bibinfo {pages} {405}
  (\bibinfo {year} {1931}{\natexlab{a}})}\BibitemShut {NoStop}%
\bibitem [{\citenamefont
  {Onsager}(1931{\natexlab{b}})}]{onsager1931reciprocal_2}%
  \BibitemOpen
  \bibfield  {author} {\bibinfo {author} {\bibfnamefont {L.}~\bibnamefont
  {Onsager}},\ }\href@noop {} {\bibfield  {journal} {\bibinfo  {journal}
  {Physical review}\ }\textbf {\bibinfo {volume} {38}},\ \bibinfo {pages}
  {2265} (\bibinfo {year} {1931}{\natexlab{b}})}\BibitemShut {NoStop}%
\bibitem [{\citenamefont {Adams}\ \emph {et~al.}(2013)\citenamefont {Adams},
  \citenamefont {Dirr}, \citenamefont {Peletier},\ and\ \citenamefont
  {Zimmer}}]{adams2013large}%
  \BibitemOpen
  \bibfield  {author} {\bibinfo {author} {\bibfnamefont {S.}~\bibnamefont
  {Adams}}, \bibinfo {author} {\bibfnamefont {N.}~\bibnamefont {Dirr}},
  \bibinfo {author} {\bibfnamefont {M.}~\bibnamefont {Peletier}}, \ and\
  \bibinfo {author} {\bibfnamefont {J.}~\bibnamefont {Zimmer}},\ }\href@noop {}
  {\bibfield  {journal} {\bibinfo  {journal} {Philosophical Transactions of the
  Royal Society A: Mathematical, Physical and Engineering Sciences}\ }\textbf
  {\bibinfo {volume} {371}},\ \bibinfo {pages} {20120341} (\bibinfo {year}
  {2013})}\BibitemShut {NoStop}%
\bibitem [{\citenamefont {Melcher}\ and\ \citenamefont
  {Taylor}(1969)}]{melcher1969electrohydrodynamics}%
  \BibitemOpen
  \bibfield  {author} {\bibinfo {author} {\bibfnamefont {J.}~\bibnamefont
  {Melcher}}\ and\ \bibinfo {author} {\bibfnamefont {G.}~\bibnamefont
  {Taylor}},\ }\href@noop {} {\bibfield  {journal} {\bibinfo  {journal} {Annual
  review of fluid mechanics}\ }\textbf {\bibinfo {volume} {1}},\ \bibinfo
  {pages} {111} (\bibinfo {year} {1969})}\BibitemShut {NoStop}%
\bibitem [{\citenamefont {Saville}(1997)}]{saville1997electrohydrodynamics}%
  \BibitemOpen
  \bibfield  {author} {\bibinfo {author} {\bibfnamefont {D.}~\bibnamefont
  {Saville}},\ }\href@noop {} {\bibfield  {journal} {\bibinfo  {journal}
  {Annual review of fluid mechanics}\ }\textbf {\bibinfo {volume} {29}},\
  \bibinfo {pages} {27} (\bibinfo {year} {1997})}\BibitemShut {NoStop}%
\bibitem [{\citenamefont {Melcher}(1981)}]{melcher1981continuum}%
  \BibitemOpen
  \bibfield  {author} {\bibinfo {author} {\bibfnamefont {J.~R.}\ \bibnamefont
  {Melcher}},\ }\href@noop {} {\emph {\bibinfo {title} {Continuum
  electromechanics}}},\ Vol.~\bibinfo {volume} {2}\ (\bibinfo  {publisher} {MIT
  press Cambridge, MA},\ \bibinfo {year} {1981})\BibitemShut {NoStop}%
\bibitem [{\citenamefont {Schnitzer}\ and\ \citenamefont
  {Yariv}(2015)}]{schnitzer2015taylor}%
  \BibitemOpen
  \bibfield  {author} {\bibinfo {author} {\bibfnamefont {O.}~\bibnamefont
  {Schnitzer}}\ and\ \bibinfo {author} {\bibfnamefont {E.}~\bibnamefont
  {Yariv}},\ }\href@noop {} {\bibfield  {journal} {\bibinfo  {journal} {Journal
  of Fluid Mechanics}\ }\textbf {\bibinfo {volume} {773}},\ \bibinfo {pages}
  {1} (\bibinfo {year} {2015})}\BibitemShut {NoStop}%
\bibitem [{\citenamefont {Bazant}(2015)}]{bazant2015electrokinetics}%
  \BibitemOpen
  \bibfield  {author} {\bibinfo {author} {\bibfnamefont {M.~Z.}\ \bibnamefont
  {Bazant}},\ }\href@noop {} {\bibfield  {journal} {\bibinfo  {journal}
  {Journal of Fluid Mechanics}\ }\textbf {\bibinfo {volume} {782}},\ \bibinfo
  {pages} {1} (\bibinfo {year} {2015})}\BibitemShut {NoStop}%
\bibitem [{\citenamefont {Kittel}(1976)}]{kittel1976introduction}%
  \BibitemOpen
  \bibfield  {author} {\bibinfo {author} {\bibfnamefont {C.}~\bibnamefont
  {Kittel}},\ }\href@noop {} {\emph {\bibinfo {title} {Introduction to solid
  state physics}}},\ Vol.~\bibinfo {volume} {8}\ (\bibinfo  {publisher} {Wiley
  New York},\ \bibinfo {year} {1976})\BibitemShut {NoStop}%
\bibitem [{\citenamefont {Vasileiadis}\ \emph {et~al.}(2018)\citenamefont
  {Vasileiadis}, \citenamefont {de~Klerk}, \citenamefont {Smith}, \citenamefont
  {Ganapathy}, \citenamefont {Harks}, \citenamefont {Bazant},\ and\
  \citenamefont {Wagemaker}}]{vasileiadis2018toward}%
  \BibitemOpen
  \bibfield  {author} {\bibinfo {author} {\bibfnamefont {A.}~\bibnamefont
  {Vasileiadis}}, \bibinfo {author} {\bibfnamefont {N.~J.}\ \bibnamefont
  {de~Klerk}}, \bibinfo {author} {\bibfnamefont {R.~B.}\ \bibnamefont {Smith}},
  \bibinfo {author} {\bibfnamefont {S.}~\bibnamefont {Ganapathy}}, \bibinfo
  {author} {\bibfnamefont {P.~P.~R.}\ \bibnamefont {Harks}}, \bibinfo {author}
  {\bibfnamefont {M.~Z.}\ \bibnamefont {Bazant}}, \ and\ \bibinfo {author}
  {\bibfnamefont {M.}~\bibnamefont {Wagemaker}},\ }\href@noop {} {\bibfield
  {journal} {\bibinfo  {journal} {Advanced Functional Materials}\ }\textbf
  {\bibinfo {volume} {28}},\ \bibinfo {pages} {1705992} (\bibinfo {year}
  {2018})}\BibitemShut {NoStop}%
\bibitem [{\citenamefont {Liu}\ \emph {et~al.}(2017)\citenamefont {Liu},
  \citenamefont {Lian}, \citenamefont {Sun}, \citenamefont {Zhao},
  \citenamefont {Shi},\ and\ \citenamefont {Song}}]{liu2017first}%
  \BibitemOpen
  \bibfield  {author} {\bibinfo {author} {\bibfnamefont {Y.}~\bibnamefont
  {Liu}}, \bibinfo {author} {\bibfnamefont {J.}~\bibnamefont {Lian}}, \bibinfo
  {author} {\bibfnamefont {Z.}~\bibnamefont {Sun}}, \bibinfo {author}
  {\bibfnamefont {M.}~\bibnamefont {Zhao}}, \bibinfo {author} {\bibfnamefont
  {Y.}~\bibnamefont {Shi}}, \ and\ \bibinfo {author} {\bibfnamefont
  {H.}~\bibnamefont {Song}},\ }\href@noop {} {\bibfield  {journal} {\bibinfo
  {journal} {Chemical Physics Letters}\ }\textbf {\bibinfo {volume} {677}},\
  \bibinfo {pages} {114} (\bibinfo {year} {2017})}\BibitemShut {NoStop}%
\bibitem [{\citenamefont {Wagemaker}\ \emph {et~al.}(2009)\citenamefont
  {Wagemaker}, \citenamefont {van Eck}, \citenamefont {Kentgens},\ and\
  \citenamefont {Mulder}}]{wagemaker2009li}%
  \BibitemOpen
  \bibfield  {author} {\bibinfo {author} {\bibfnamefont {M.}~\bibnamefont
  {Wagemaker}}, \bibinfo {author} {\bibfnamefont {E.~R.}\ \bibnamefont {van
  Eck}}, \bibinfo {author} {\bibfnamefont {A.~P.}\ \bibnamefont {Kentgens}}, \
  and\ \bibinfo {author} {\bibfnamefont {F.~M.}\ \bibnamefont {Mulder}},\
  }\href@noop {} {\bibfield  {journal} {\bibinfo  {journal} {The Journal of
  Physical Chemistry B}\ }\textbf {\bibinfo {volume} {113}},\ \bibinfo {pages}
  {224} (\bibinfo {year} {2009})}\BibitemShut {NoStop}%
\bibitem [{\citenamefont {Schmidt}\ \emph {et~al.}(2015)\citenamefont
  {Schmidt}, \citenamefont {Bottke}, \citenamefont {Sternad}, \citenamefont
  {Gollob}, \citenamefont {Hennige},\ and\ \citenamefont
  {Wilkening}}]{schmidt2015small}%
  \BibitemOpen
  \bibfield  {author} {\bibinfo {author} {\bibfnamefont {W.}~\bibnamefont
  {Schmidt}}, \bibinfo {author} {\bibfnamefont {P.}~\bibnamefont {Bottke}},
  \bibinfo {author} {\bibfnamefont {M.}~\bibnamefont {Sternad}}, \bibinfo
  {author} {\bibfnamefont {P.}~\bibnamefont {Gollob}}, \bibinfo {author}
  {\bibfnamefont {V.}~\bibnamefont {Hennige}}, \ and\ \bibinfo {author}
  {\bibfnamefont {M.}~\bibnamefont {Wilkening}},\ }\href@noop {} {\bibfield
  {journal} {\bibinfo  {journal} {Chemistry of Materials}\ }\textbf {\bibinfo
  {volume} {27}},\ \bibinfo {pages} {1740} (\bibinfo {year}
  {2015})}\BibitemShut {NoStop}%
\bibitem [{\citenamefont {Wilkening}\ \emph {et~al.}(2007)\citenamefont
  {Wilkening}, \citenamefont {Iwaniak}, \citenamefont {Heine}, \citenamefont
  {Epp}, \citenamefont {Kleinert}, \citenamefont {Behrens}, \citenamefont
  {Nuspl}, \citenamefont {Bensch},\ and\ \citenamefont
  {Heitjans}}]{wilkening2007microscopic}%
  \BibitemOpen
  \bibfield  {author} {\bibinfo {author} {\bibfnamefont {M.}~\bibnamefont
  {Wilkening}}, \bibinfo {author} {\bibfnamefont {W.}~\bibnamefont {Iwaniak}},
  \bibinfo {author} {\bibfnamefont {J.}~\bibnamefont {Heine}}, \bibinfo
  {author} {\bibfnamefont {V.}~\bibnamefont {Epp}}, \bibinfo {author}
  {\bibfnamefont {A.}~\bibnamefont {Kleinert}}, \bibinfo {author}
  {\bibfnamefont {M.}~\bibnamefont {Behrens}}, \bibinfo {author} {\bibfnamefont
  {G.}~\bibnamefont {Nuspl}}, \bibinfo {author} {\bibfnamefont
  {W.}~\bibnamefont {Bensch}}, \ and\ \bibinfo {author} {\bibfnamefont
  {P.}~\bibnamefont {Heitjans}},\ }\href@noop {} {\bibfield  {journal}
  {\bibinfo  {journal} {Physical Chemistry Chemical Physics}\ }\textbf
  {\bibinfo {volume} {9}},\ \bibinfo {pages} {6199} (\bibinfo {year}
  {2007})}\BibitemShut {NoStop}%
\bibitem [{\citenamefont {Bazant}\ \emph {et~al.}(2004)\citenamefont {Bazant},
  \citenamefont {Thornton},\ and\ \citenamefont {Ajdari}}]{bazant2004diffuse}%
  \BibitemOpen
  \bibfield  {author} {\bibinfo {author} {\bibfnamefont {M.~Z.}\ \bibnamefont
  {Bazant}}, \bibinfo {author} {\bibfnamefont {K.}~\bibnamefont {Thornton}}, \
  and\ \bibinfo {author} {\bibfnamefont {A.}~\bibnamefont {Ajdari}},\
  }\href@noop {} {\bibfield  {journal} {\bibinfo  {journal} {Physical review
  E}\ }\textbf {\bibinfo {volume} {70}},\ \bibinfo {pages} {021506} (\bibinfo
  {year} {2004})}\BibitemShut {NoStop}%
\bibitem [{\citenamefont {Fraggedakis}\ \emph {et~al.}(2017)\citenamefont
  {Fraggedakis}, \citenamefont {Papaioannou}, \citenamefont {Dimakopoulos},\
  and\ \citenamefont {Tsamopoulos}}]{fraggedakis2017discretization}%
  \BibitemOpen
  \bibfield  {author} {\bibinfo {author} {\bibfnamefont {D.}~\bibnamefont
  {Fraggedakis}}, \bibinfo {author} {\bibfnamefont {J.}~\bibnamefont
  {Papaioannou}}, \bibinfo {author} {\bibfnamefont {Y.}~\bibnamefont
  {Dimakopoulos}}, \ and\ \bibinfo {author} {\bibfnamefont {J.}~\bibnamefont
  {Tsamopoulos}},\ }\href@noop {} {\bibfield  {journal} {\bibinfo  {journal}
  {Journal of Computational Physics}\ }\textbf {\bibinfo {volume} {344}},\
  \bibinfo {pages} {127} (\bibinfo {year} {2017})}\BibitemShut {NoStop}%
\bibitem [{\citenamefont {Fraggedakis}\ \emph {et~al.}(2015)\citenamefont
  {Fraggedakis}, \citenamefont {Kouris}, \citenamefont {Dimakopoulos},\ and\
  \citenamefont {Tsamopoulos}}]{fraggedakis2015flow}%
  \BibitemOpen
  \bibfield  {author} {\bibinfo {author} {\bibfnamefont {D.}~\bibnamefont
  {Fraggedakis}}, \bibinfo {author} {\bibfnamefont {C.}~\bibnamefont {Kouris}},
  \bibinfo {author} {\bibfnamefont {Y.}~\bibnamefont {Dimakopoulos}}, \ and\
  \bibinfo {author} {\bibfnamefont {J.}~\bibnamefont {Tsamopoulos}},\
  }\href@noop {} {\bibfield  {journal} {\bibinfo  {journal} {Physics of
  Fluids}\ }\textbf {\bibinfo {volume} {27}},\ \bibinfo {pages} {082102}
  (\bibinfo {year} {2015})}\BibitemShut {NoStop}%
\bibitem [{\citenamefont {Leal}(2007)}]{leal2007advanced}%
  \BibitemOpen
  \bibfield  {author} {\bibinfo {author} {\bibfnamefont {L.~G.}\ \bibnamefont
  {Leal}},\ }\href@noop {} {\emph {\bibinfo {title} {Advanced transport
  phenomena: fluid mechanics and convective transport processes}}},\
  Vol.~\bibinfo {volume} {7}\ (\bibinfo  {publisher} {Cambridge University
  Press},\ \bibinfo {year} {2007})\BibitemShut {NoStop}%
\bibitem [{\citenamefont {Voorhees}(1985)}]{voorhees1985theory}%
  \BibitemOpen
  \bibfield  {author} {\bibinfo {author} {\bibfnamefont {P.~W.}\ \bibnamefont
  {Voorhees}},\ }\href@noop {} {\bibfield  {journal} {\bibinfo  {journal}
  {Journal of Statistical Physics}\ }\textbf {\bibinfo {volume} {38}},\
  \bibinfo {pages} {231} (\bibinfo {year} {1985})}\BibitemShut {NoStop}%
\bibitem [{\citenamefont {Taylor}(1964)}]{taylor1964disintegration}%
  \BibitemOpen
  \bibfield  {author} {\bibinfo {author} {\bibfnamefont {G.~I.}\ \bibnamefont
  {Taylor}},\ }\href@noop {} {\bibfield  {journal} {\bibinfo  {journal}
  {Proceedings of the Royal Society of London. Series A. Mathematical and
  Physical Sciences}\ }\textbf {\bibinfo {volume} {280}},\ \bibinfo {pages}
  {383} (\bibinfo {year} {1964})}\BibitemShut {NoStop}%
\bibitem [{\citenamefont {Zhou}\ and\ \citenamefont
  {Ramanathan}(2015)}]{Zhou2015}%
  \BibitemOpen
  \bibfield  {author} {\bibinfo {author} {\bibfnamefont {Y.}~\bibnamefont
  {Zhou}}\ and\ \bibinfo {author} {\bibfnamefont {S.}~\bibnamefont
  {Ramanathan}},\ }\href {\doibase 10.1109/JPROC.2015.2431914} {\bibfield
  {journal} {\bibinfo  {journal} {Proceedings of the IEEE}\ }\textbf {\bibinfo
  {volume} {103}},\ \bibinfo {pages} {1289} (\bibinfo {year}
  {2015})}\BibitemShut {NoStop}%
\bibitem [{\citenamefont {Bisri}\ \emph {et~al.}(2017)\citenamefont {Bisri},
  \citenamefont {Shimizu}, \citenamefont {Nakano},\ and\ \citenamefont
  {Iwasa}}]{Bisri2017}%
  \BibitemOpen
  \bibfield  {author} {\bibinfo {author} {\bibfnamefont {S.~Z.}\ \bibnamefont
  {Bisri}}, \bibinfo {author} {\bibfnamefont {S.}~\bibnamefont {Shimizu}},
  \bibinfo {author} {\bibfnamefont {M.}~\bibnamefont {Nakano}}, \ and\ \bibinfo
  {author} {\bibfnamefont {Y.}~\bibnamefont {Iwasa}},\ }\href {\doibase
  10.1002/adma.201607054} {\bibfield  {journal} {\bibinfo  {journal} {Advanced
  Materials}\ }\textbf {\bibinfo {volume} {29}},\ \bibinfo {pages} {1}
  (\bibinfo {year} {2017})}\BibitemShut {NoStop}%
\bibitem [{\citenamefont {Lee}\ \emph {et~al.}(2015)\citenamefont {Lee},
  \citenamefont {Lee},\ and\ \citenamefont {Noh}}]{Lee2015}%
  \BibitemOpen
  \bibfield  {author} {\bibinfo {author} {\bibfnamefont {J.~S.}\ \bibnamefont
  {Lee}}, \bibinfo {author} {\bibfnamefont {S.}~\bibnamefont {Lee}}, \ and\
  \bibinfo {author} {\bibfnamefont {T.~W.}\ \bibnamefont {Noh}},\ }\href
  {\doibase 10.1063/1.4929512} {\bibfield  {journal} {\bibinfo  {journal}
  {Applied Physics Reviews}\ }\textbf {\bibinfo {volume} {2}} (\bibinfo {year}
  {2015}),\ 10.1063/1.4929512}\BibitemShut {NoStop}%
\bibitem [{\citenamefont {Wang}\ \emph {et~al.}(2020)\citenamefont {Wang},
  \citenamefont {Wu}, \citenamefont {Burr}, \citenamefont {Hwang},
  \citenamefont {Wang}, \citenamefont {Xia},\ and\ \citenamefont
  {Yang}}]{Wang2020}%
  \BibitemOpen
  \bibfield  {author} {\bibinfo {author} {\bibfnamefont {Z.}~\bibnamefont
  {Wang}}, \bibinfo {author} {\bibfnamefont {H.}~\bibnamefont {Wu}}, \bibinfo
  {author} {\bibfnamefont {G.~W.}\ \bibnamefont {Burr}}, \bibinfo {author}
  {\bibfnamefont {C.~S.}\ \bibnamefont {Hwang}}, \bibinfo {author}
  {\bibfnamefont {K.~L.}\ \bibnamefont {Wang}}, \bibinfo {author}
  {\bibfnamefont {Q.}~\bibnamefont {Xia}}, \ and\ \bibinfo {author}
  {\bibfnamefont {J.~J.}\ \bibnamefont {Yang}},\ }\href {\doibase
  10.1038/s41578-019-0159-3} {\bibfield  {journal} {\bibinfo  {journal} {Nature
  Reviews Materials}\ } (\bibinfo {year} {2020}),\
  10.1038/s41578-019-0159-3}\BibitemShut {NoStop}%
\bibitem [{\citenamefont {Sun}\ \emph {et~al.}(2019)\citenamefont {Sun},
  \citenamefont {Gao}, \citenamefont {Chi}, \citenamefont {Xia}, \citenamefont
  {Yang}, \citenamefont {Qian},\ and\ \citenamefont {Wu}}]{Sun2019}%
  \BibitemOpen
  \bibfield  {author} {\bibinfo {author} {\bibfnamefont {W.}~\bibnamefont
  {Sun}}, \bibinfo {author} {\bibfnamefont {B.}~\bibnamefont {Gao}}, \bibinfo
  {author} {\bibfnamefont {M.}~\bibnamefont {Chi}}, \bibinfo {author}
  {\bibfnamefont {Q.}~\bibnamefont {Xia}}, \bibinfo {author} {\bibfnamefont
  {J.~J.}\ \bibnamefont {Yang}}, \bibinfo {author} {\bibfnamefont
  {H.}~\bibnamefont {Qian}}, \ and\ \bibinfo {author} {\bibfnamefont
  {H.}~\bibnamefont {Wu}},\ }\href {\doibase 10.1038/s41467-019-11411-6}
  {\bibfield  {journal} {\bibinfo  {journal} {Nature Communications}\ }\textbf
  {\bibinfo {volume} {10}},\ \bibinfo {pages} {1} (\bibinfo {year}
  {2019})}\BibitemShut {NoStop}%
\bibitem [{\citenamefont {Green}\ and\ \citenamefont
  {Aimone}(2019)}]{Green1928}%
  \BibitemOpen
  \bibfield  {author} {\bibinfo {author} {\bibfnamefont {S.}~\bibnamefont
  {Green}}\ and\ \bibinfo {author} {\bibfnamefont {J.~B.}\ \bibnamefont
  {Aimone}},\ }\href {\doibase 10.1038/s41928-019-0224-3} {\bibfield  {journal}
  {\bibinfo  {journal} {Nature Electronics}\ }\textbf {\bibinfo {volume} {2}},\
  \bibinfo {pages} {96} (\bibinfo {year} {2019})}\BibitemShut {NoStop}%
\bibitem [{\citenamefont {Luntz}\ \emph {et~al.}(2015)\citenamefont {Luntz},
  \citenamefont {Voss},\ and\ \citenamefont {Reuter}}]{luntz2015interfacial}%
  \BibitemOpen
  \bibfield  {author} {\bibinfo {author} {\bibfnamefont {A.~C.}\ \bibnamefont
  {Luntz}}, \bibinfo {author} {\bibfnamefont {J.}~\bibnamefont {Voss}}, \ and\
  \bibinfo {author} {\bibfnamefont {K.}~\bibnamefont {Reuter}},\ }\href@noop {}
  {\enquote {\bibinfo {title} {Interfacial challenges in solid-state li ion
  batteries},}\ } (\bibinfo {year} {2015})\BibitemShut {NoStop}%
\bibitem [{\citenamefont {Maxisch}\ \emph {et~al.}(2006)\citenamefont
  {Maxisch}, \citenamefont {Zhou},\ and\ \citenamefont
  {Ceder}}]{maxisch2006ab}%
  \BibitemOpen
  \bibfield  {author} {\bibinfo {author} {\bibfnamefont {T.}~\bibnamefont
  {Maxisch}}, \bibinfo {author} {\bibfnamefont {F.}~\bibnamefont {Zhou}}, \
  and\ \bibinfo {author} {\bibfnamefont {G.}~\bibnamefont {Ceder}},\
  }\href@noop {} {\bibfield  {journal} {\bibinfo  {journal} {Physical review
  B}\ }\textbf {\bibinfo {volume} {73}},\ \bibinfo {pages} {104301} (\bibinfo
  {year} {2006})}\BibitemShut {NoStop}%
\bibitem [{\citenamefont {Malik}\ \emph {et~al.}(2013)\citenamefont {Malik},
  \citenamefont {Abdellahi},\ and\ \citenamefont {Ceder}}]{malik2013critical}%
  \BibitemOpen
  \bibfield  {author} {\bibinfo {author} {\bibfnamefont {R.}~\bibnamefont
  {Malik}}, \bibinfo {author} {\bibfnamefont {A.}~\bibnamefont {Abdellahi}}, \
  and\ \bibinfo {author} {\bibfnamefont {G.}~\bibnamefont {Ceder}},\
  }\href@noop {} {\bibfield  {journal} {\bibinfo  {journal} {Journal of the
  electrochemical society}\ }\textbf {\bibinfo {volume} {160}},\ \bibinfo
  {pages} {A3179} (\bibinfo {year} {2013})}\BibitemShut {NoStop}%
\bibitem [{\citenamefont {Kim}\ \emph {et~al.}(2011)\citenamefont {Kim},
  \citenamefont {Jeong},\ and\ \citenamefont {Hwang}}]{kim2011nanofilamentary}%
  \BibitemOpen
  \bibfield  {author} {\bibinfo {author} {\bibfnamefont {K.~M.}\ \bibnamefont
  {Kim}}, \bibinfo {author} {\bibfnamefont {D.~S.}\ \bibnamefont {Jeong}}, \
  and\ \bibinfo {author} {\bibfnamefont {C.~S.}\ \bibnamefont {Hwang}},\
  }\href@noop {} {\bibfield  {journal} {\bibinfo  {journal} {Nanotechnology}\
  }\textbf {\bibinfo {volume} {22}},\ \bibinfo {pages} {254002} (\bibinfo
  {year} {2011})}\BibitemShut {NoStop}%
\bibitem [{\citenamefont {Ielmini}(2011)}]{ielmini2011modeling}%
  \BibitemOpen
  \bibfield  {author} {\bibinfo {author} {\bibfnamefont {D.}~\bibnamefont
  {Ielmini}},\ }\href@noop {} {\bibfield  {journal} {\bibinfo  {journal} {IEEE
  Transactions on Electron Devices}\ }\textbf {\bibinfo {volume} {58}},\
  \bibinfo {pages} {4309} (\bibinfo {year} {2011})}\BibitemShut {NoStop}%
\bibitem [{\citenamefont {Strachan}\ \emph {et~al.}(2011)\citenamefont
  {Strachan}, \citenamefont {Strukov}, \citenamefont {Borghetti}, \citenamefont
  {Yang}, \citenamefont {Medeiros-Ribeiro},\ and\ \citenamefont
  {Williams}}]{strachan2011switching}%
  \BibitemOpen
  \bibfield  {author} {\bibinfo {author} {\bibfnamefont {J.~P.}\ \bibnamefont
  {Strachan}}, \bibinfo {author} {\bibfnamefont {D.~B.}\ \bibnamefont
  {Strukov}}, \bibinfo {author} {\bibfnamefont {J.}~\bibnamefont {Borghetti}},
  \bibinfo {author} {\bibfnamefont {J.~J.}\ \bibnamefont {Yang}}, \bibinfo
  {author} {\bibfnamefont {G.}~\bibnamefont {Medeiros-Ribeiro}}, \ and\
  \bibinfo {author} {\bibfnamefont {R.~S.}\ \bibnamefont {Williams}},\
  }\href@noop {} {\bibfield  {journal} {\bibinfo  {journal} {Nanotechnology}\
  }\textbf {\bibinfo {volume} {22}},\ \bibinfo {pages} {254015} (\bibinfo
  {year} {2011})}\BibitemShut {NoStop}%
\bibitem [{\citenamefont {Pickett}\ \emph {et~al.}(2013)\citenamefont
  {Pickett}, \citenamefont {Medeiros-Ribeiro},\ and\ \citenamefont
  {Williams}}]{pickett2013scalable}%
  \BibitemOpen
  \bibfield  {author} {\bibinfo {author} {\bibfnamefont {M.~D.}\ \bibnamefont
  {Pickett}}, \bibinfo {author} {\bibfnamefont {G.}~\bibnamefont
  {Medeiros-Ribeiro}}, \ and\ \bibinfo {author} {\bibfnamefont {R.~S.}\
  \bibnamefont {Williams}},\ }\href@noop {} {\bibfield  {journal} {\bibinfo
  {journal} {Nature materials}\ }\textbf {\bibinfo {volume} {12}},\ \bibinfo
  {pages} {114} (\bibinfo {year} {2013})}\BibitemShut {NoStop}%
\bibitem [{\citenamefont {Wang}\ \emph
  {et~al.}(2019{\natexlab{a}})\citenamefont {Wang}, \citenamefont {Laudato},
  \citenamefont {Ambrosi}, \citenamefont {Bricalli}, \citenamefont {Covi},
  \citenamefont {Lin},\ and\ \citenamefont {Ielmini}}]{Wang2019a}%
  \BibitemOpen
  \bibfield  {author} {\bibinfo {author} {\bibfnamefont {W.}~\bibnamefont
  {Wang}}, \bibinfo {author} {\bibfnamefont {M.}~\bibnamefont {Laudato}},
  \bibinfo {author} {\bibfnamefont {E.}~\bibnamefont {Ambrosi}}, \bibinfo
  {author} {\bibfnamefont {A.}~\bibnamefont {Bricalli}}, \bibinfo {author}
  {\bibfnamefont {E.}~\bibnamefont {Covi}}, \bibinfo {author} {\bibfnamefont
  {Y.~H.}\ \bibnamefont {Lin}}, \ and\ \bibinfo {author} {\bibfnamefont
  {D.}~\bibnamefont {Ielmini}},\ }\href {\doibase 10.1109/TED.2019.2928890}
  {\bibfield  {journal} {\bibinfo  {journal} {IEEE Transactions on Electron
  Devices}\ }\textbf {\bibinfo {volume} {66}},\ \bibinfo {pages} {3795}
  (\bibinfo {year} {2019}{\natexlab{a}})}\BibitemShut {NoStop}%
\bibitem [{\citenamefont {Wang}\ \emph
  {et~al.}(2019{\natexlab{b}})\citenamefont {Wang}, \citenamefont {Laudato},
  \citenamefont {Ambrosi}, \citenamefont {Bricalli}, \citenamefont {Covi},
  \citenamefont {Lin},\ and\ \citenamefont {Ielmini}}]{Wang2019b}%
  \BibitemOpen
  \bibfield  {author} {\bibinfo {author} {\bibfnamefont {W.}~\bibnamefont
  {Wang}}, \bibinfo {author} {\bibfnamefont {M.}~\bibnamefont {Laudato}},
  \bibinfo {author} {\bibfnamefont {E.}~\bibnamefont {Ambrosi}}, \bibinfo
  {author} {\bibfnamefont {A.}~\bibnamefont {Bricalli}}, \bibinfo {author}
  {\bibfnamefont {E.}~\bibnamefont {Covi}}, \bibinfo {author} {\bibfnamefont
  {Y.~H.}\ \bibnamefont {Lin}}, \ and\ \bibinfo {author} {\bibfnamefont
  {D.}~\bibnamefont {Ielmini}},\ }\href {\doibase 10.1109/TED.2019.2928888}
  {\bibfield  {journal} {\bibinfo  {journal} {IEEE Transactions on Electron
  Devices}\ }\textbf {\bibinfo {volume} {66}},\ \bibinfo {pages} {3802}
  (\bibinfo {year} {2019}{\natexlab{b}})}\BibitemShut {NoStop}%
\bibitem [{\citenamefont {Cogswell}\ and\ \citenamefont
  {Bazant}(2012)}]{cogswell2012coherency}%
  \BibitemOpen
  \bibfield  {author} {\bibinfo {author} {\bibfnamefont {D.~A.}\ \bibnamefont
  {Cogswell}}\ and\ \bibinfo {author} {\bibfnamefont {M.~Z.}\ \bibnamefont
  {Bazant}},\ }\href@noop {} {\bibfield  {journal} {\bibinfo  {journal} {ACS
  nano}\ }\textbf {\bibinfo {volume} {6}},\ \bibinfo {pages} {2215} (\bibinfo
  {year} {2012})}\BibitemShut {NoStop}%
\end{thebibliography}%

\newpage

% \newpage

\end{document}